# A rich molecular chemistry in the gas of the IC 348 star cluster of the Perseus Molecular Cloud

Susana Iglesias-Groth[1,2] & Martina Marin-Dobrincic[3]

[1]*Instituto de Astrofísica de Canarias*

C/ Via Láctea s/n, 38200 La Laguna, Tenerife, Spain

[2]*Departamento de Astrofísica. Universidad de la Laguna*

C/ Francisco Sánchez, 38200 La Laguna, Tenerife

[3]*Universidad Polictécnica de Cartagena*

C/ Campus Muralla del Mar, 30202 Cartagena, Murcia, Spain

## Abstract

We present Spitzer 10-34 $\mu$m spectroscopic observations of the diffuse gas in the inner region of the star-forming region IC 348 of the Perseus Molecular Cloud. We find evidence for the strongest mid-IR bands of common molecules as $H_2$, OH, $H_2O$, $CO_2$ and $NH_3$ and of several carbonaceous molecules which may play an important role in the production of more complex hydrocarbons: HCN, $C_2H_2$, $C_4H_2$, $HC_3N$, $HC_5N$, $C_2H_6$, $C_6H_2$, $C_6H_6$. The excitation diagram of $H_2$ reveals the presence of warm gas (270 +- 30 K) at the observed locations. Assuming this temperature, the derived abundances of $CO_2$ and $NH_3$ relative to $H_2$ are $10^{-8}$ and $10^{-7}$, respectively. From the water lines we obtain an abundance of order $10^{-6}$ and higher gas temperatures. The abundances derived for HCN and $C_2H_2$, key molecules in the development of prebiotic building blocks, are of order $10^{-7}$ and $10^{-9}$, respectively. More complex molecules such as PAHs and the fullerenes $C_{60}$ and $C_{70}$ are also present. IC 348 appears to be very rich and diverse in molecular content. The JWST spectroscopic capabilities may provide details on the spatial distribution of all these molecules and extend the present search to more complex hydrocarbons.

# 1. INTRODUCTION

The young stellar cluster IC 348 (age ≈ 2 Myr), located at the eastern end of the well-known Perseus Molecular Cloud complex (Herbig 1998, Luhman et al. 2003, Bally et al 2008, Luhman et al. 2016) at a distance of 321 ±10 pc (Ortiz-León et al. 2018), is one of the nearest star-forming regions. The Perseus Molecular Cloud provided one of the first unambiguous detections of the anomalous microwave emission (Watson et al. 2005; Planck collaboration 2011), radiation probably caused by electric dipole emission (Draine and Lazarian 1998) from a diffuse distribution of fast spinning carbonaceous molecules (polycyclic aromatic hydrocarbons, hydrogenated fullerenes, etc.). Revealing the presence of individual species of carbonaceous molecules in this star-forming region may provide insight on the mechanism responsible of this microwave emission.

IC 348 has already shown the presence of complex carbon-based molecules such as fullerenes and PAHs (see Iglesias-Groth 2019), but it remains unclear if the observed fullerenes were formed in situ or originated somewhere else and subsequently contaminated the interstellar gas in this star-forming region (Murga et al. 2022). It is interesting to explore the presence of other carbonaceous molecules that are known to intervene in the production of hydrocarbons as their detection may provide some insight into the chemical networks operating.

A few dozen molecules have been detected in the interstellar medium of other star-forming regions (Cazaux et al. 2003, Jorgensen et al. 2012, Kahane et al. 2013, Favrè et al. 2018) and in protoplanetary disks (Dutrey et al. 2014). These detections include aldehydes, acids, ketones, and sugars. The simplest organic acid, formic acid (HCOOH), which contains the carboxyl group, one of the main functional groups of amino acids, has been detected in low-mass star-forming regions (Lefloch et al. 2017) and in a protoplanetary disk (Favrè et al. 2018). Many of these molecules are known to present bands in the mid-IR which can be investigated using available spectroscopic observations of the *Spitzer Space Telescope*.

In this paper, using moderate spectral resolution data from *Spitzer Space Telescope* we report evidence for the presence in the IC 348 interstellar gas of many carbonaceous molecules commonly reported in proto-planetary disks and young stellar objects.

## 2. OBSERVATIONS

We have used high spectral resolution (R≈600) archive spectra obtained with the Infrared Spectrograph (IRS) onboard the *Spitzer Space Telescope* at various interstellar medium locations in the central region of IC 348 in the Perseus Molecular Cloud. All the selected observations were located within 10 arcmin of the most luminous star of the cluster, HD 281159, within the region marked in Fig. 1. Fully reduced spectra acquired with the Short-High (S-H, 9.8-19.5 $\mu$m) and Long-High (L-H, 19.5-36 $\mu$m) modules were taken from the Combined Atlas of Spitzer/IRS Sources (CASSIS; http:// cassis.sirtf.com, Lebouteiller et al. 2015) which provides reduced and flux calibrated data. The spectra used from both modules of the IRS are identified with the corresponding AOR numbers in Table 1. Coordinates of the ISM pointings are also listed. The present search is focused on interstellar locations so we have selected pointings which in the original observing programmes, mainly focused on the study of protoplanetry discs, served as background observations for nearby targets.

The S-H and L-H spectra are not available for the same interstellar locations; however, we do not intend to carry out studies of individual locations, on the contrary, we averaged the available spectra from different pointings in the core region of IC 348 from both IRS modules in order to gain as high signal to noise as possible, so as to facilitate the search for weak bands.

## 3. DATA REDUCTION

The CASSIS pipeline takes the Basic Calibrated Data (BCD) images, the associated uncertainty images and the bad pixel mask as starting products. The BCD images are produced by the Spitzer Science Center BCD pipeline. The High-resolution modules of the IRS are significantly affected by cosmic ray hits and spurious features, thus the CASSIS pipeline (Lebouteiller et al. 2015) pays special attention to the exposure combination and image cleaning to remove any bad pixels in order to minimize these effects. This pipeline incorporates various improvements with respect to Lebouteiller et al. (2011) including several techniques for optimal extraction and a differential method that eliminates low-level rogue pixels.

As we are interested in observations of the general diffuse interstellar medium, we will use the CASSIS full-aperture extraction for extended sources. In order to compute fluxes, this extraction method co-adds the pixels in the detector area corresponding to one wavelength value. Since the flux is integrated, the presence of bad pixels anywhere within this area is especially damaging and bad pixels need to be picked out and replaced. The "cleaning" is carried out on the combined image of all exposures obtained for a given target. Bad pixel fluxes are substituted using neighbours whose flux is more reliable when exposures have been combined. The most problematic are the rogue pixels which have elevated dark current that changes unpredictably with time. The CASSIS pipeline uses IRSCLEAN, (http://irsa.ipac.caltech.edu/data/SPITZER/docs/dataanalysistools/tools/ irsclean) an IDL tool for creating bad pixel masks from Spitzer IRS BCD (and pre-BCD) image data, and "cleaning" the masked pixels in a set of data to substitute bad pixels. As described by Lebouteiller et al. (2015) this includes pixels with no values (NaNs), with a high bad pixel mask value (>256), bad pixels and rogue pixels flagged in the campaign mask, pixels with a large uncertainty and negative pixels.

The new pixel values calculated by the IRSCLEAN algorithm are mainly based on neighbouring pixels taking into account uncertainties and mask values propagated for each step. Subsequently, the full-aperture extraction is performed using the standard tool in SMART. An error on the flux is ultimately calculated using the quadratic sum of the pixel uncertainties in the detector area corresponding to each wavelength value. No background subtraction is applied for full-aperture extraction; however, this is not expected to introduce any significant uncertainty in the measurement of wavelengths and fluxes of the relatively weak bands which we aim to detect.

As a check of the effectiveness of the bad pixel correction used in the full aperture extraction technique, we have compared the CASSIS full aperture and optimal differential extracted spectra for the well studied source RNO 90 (Pontoppidan et al 2010, Salyk et al. 2011). Figure 2a shows the spectra obtained by these two CASSIS extraction methods, small flux differences (typically less than 0.05 Jy) are seen in the continuum, probably due to the background correction applied by the optimal differential extraction. The corrections for bad pixels that both extraction techniques use do not seem to produce residual artifacts which could be confused with weak lines. If this happens, such artificial lines have peak-fluxes significantly below 0.05 Jy. As a further test we have measured fluxes for well known water bands in the full aperture spectrum of RNO 90 and compared them (see Table 2) with those reported by Blevins et al. (2016) from a completely independent reduction of the Spitzer RNO 90 spectrum. The comparison of band fluxes shows good agreement, with differences at the level of only a few percent of the measured fluxes.

The individual CASSIS full-aperture extraction of the available S-H and L-H spectra in the core of IC 348 showed very similar features and were subsequently averaged to produce a high signal to noise spectrum designated hereafter as "combined IC 348 ISM". This spectrum is displayed in Fig. 2b. Details of this spectrum will be displayed (zoomed) in many subsqunt figures when discussing the identification of molecular lines.

Line fluxes were derived from the combined IC 348 ISM spectrum by fitting a Gaussian profile and integrating the flux above a local continuum using the IRAF SPLOT routine. A polynomial of order 1 was used to fit the local continuum for a given line. Error bars were estimated using the standard deviation of the residuals of the fit. Upper-limits were derived by measuring the flux of a Gaussian with a height three times the local rms of the continuum, and with a FWHM equal to the instrumental resolution (approx. 0.02 $\mu$m). In those cases, where the rms of the continuum was difficult to establish because of neighbouring contaminant lines, the rms was taken as that of the nearest region with a "clean continuum", selected among the following spectral ranges: 10.10-10.15 $\mu$m, 11.75-11.80 $\mu$m, 14.35-14.5 $\mu$m, 14.60-14.7 $\mu$m, 15.45-15.50 $\mu$m, 16.80-16.90 $\mu$m, 20.40-20.70 $\mu$m and 27.40-27.60 $\mu$m. These regions can be seen in Fig. 2b. The minimum flux level for a line detection in the combined IC 348 ISM spectrum resulted of order 1 x $10^{-18}$ Wm$^{-2}$. Line detectability at such low fluxes opens the possibility of identifying weak transitions, enabling, in particular, a search for mid-IR transitions of previously undetected molecules.

## 4. IDENTIFICATION OF ATOMIC AND MOLECULAR LINES

McGuire (2022) lists the 241 individual molecular species detected throughout the electromagnetic spectrum in the interstellar medium. In particular, 25 individual molecular species are detected in protoplanetary disks observed in diverse star-forming regions. These molecules could already exist in the gas of the star-forming regions where the protoplanetary disks formed. Here, we present evidence for a large number (>18) of these of molecules in the mid-IR spectrum (10-34 $\mu$m) of the interstellar gas of IC 348 giving support to the idea that they may have contributed to the initial chemical composition of protoplanetary disks.

The HITRAN molecular spectroscopic database (see Gordon et al. 2022 for a most recent description) was adopted to explore the presence of such molecular lines in the observed IC 348 spectrum. The HITRAN Application Programming Interface (HAPI, Kochanov et al. 2016) was used for downloading and selecting spectroscopic transitions for each molecular specie in the HITRAN online web server and for calculating the corresponding absorption coefficients and absorption spectra in the spectral range of interest based on the available line-by-line spectroscopic parameters. The computed spectra, as well as molecular mid-IR lines and bands reported in the literature served as a reference for their identification in the spectrum of the interstellar gas in IC 348. In addition, mid-IR transitions of several atomic species commonly detected in interstellar environments were also considered.

## *4.1 Molecular hydrogen emission, $H_2$*

The IC 348 intermediate resolution IRS spectra allowed us to detect the rotational transition lines of molecular hydrogen, $H_2$ S(0) (v=0–0, J=2–0, 28.22 μm) and S(1) (v=0–0, J=3–1, 17.04 μm) lines. These lines are detected with > 10 σ amplitude sensitivity. At shorter wavelengths the upper level transition S(2), (S(2): v=0–0, J=4–2, 12.28 μm) is also detected. The v=1–1 S(3) transition at 10.18 $\mu$m could also be present in the spectrum, but with peak flux 0.007 Jy relative to the local continuum, the presence of this line is far less reliable. Fig. 3 shows the best linefits for each of the three well detected (0-0) $H_2$ lines. The plot shows the Gaussian fit after subtraction of the local continuum to each line. These transitions are strong enough to provide good flux determinations. Flux errors were determined as mentioned above.

The quadrupole moment of the pure rotational 0–0 transitions of $H_2$ makes them direct tracers of the $H_2$ gas, these transitions are excited at higher temperatures than those typically prevalent in molecular cloud interiors and trace most of the warm molecular gas with temperatures between 100 and 1000 K. The populations of these levels are thermalized under most conditions and the S(0) and S(1) lines may provide a direct measure of the masss and temperature of the bulk of warm molecular gas at T = 50-200 K. The higher pure rotational lines, as well as the vibration – rotation lines at 2 $\mu$m, are probes of the photon or shock –heated gas (Draine and Bertoldi 1999).

### 4.1.1 $H_2$ column density and excitation temperature

The excitation diagram of $H_2$ rotational emission can be used to derive the level populations and excitation temperature of warm molecular hydrogen gas (e.g. Naslim et al. 2015). Using the measured line intensities $I_{obs(u,l)}$ of each transition, the $H_2$ column densities of the upper levels ($N_u$, given in Table 3) are derived from:

$$N_u = 4\pi \lambda\, I_{obs(u,l)} / (hc\, A_{ul}), \quad (1)$$

where $A_{ul}$ is the Einstein coefficient for total radiative decay probability of the upper level. It is assumed that the lines are optically thin, the radiation is isotropic and the rotational levels of $H_2$ are thermalized. Under these assumptions, the level populations follow the Boltzmann distribution law at a given temperature and the total H2 column densities $N(H_{2tot})$ can be determined in the local thermodynamic equilibrium (LTE) condition using the formula:

$$N_u/g_u = [N(H_{2tot})/Z(T)] \exp{-E_u/k_B T}, \quad (2)$$

where $Z(T) \approx 0.0247\, T\, /[1 - \exp(-6000K/T)]$ is the partition function (Herbst et al. 1996). The statistical weight is $g_u = (2s + 1)(2j + 1)$, with spin number $s = 0$ for even J (para-$H_2$) and $s = 1$ for odd J (ortho-$H_2$) (Rosenthal, Bertoldi & Drapatz 2000).

If the $H_2$ level populations are completely dominated by collisional excitation and de-excitation then the excitation temperature equals the kinetic temperature of the gas. Population diagrams of ln ($N_u/g_u$) versus $E_u/k_B T$ allow investigation of whether the levels are thermalized. In this case, the upper-level column densities normalized with the statistical weight, $g_u$, on a logarithmic scale as a function of the upper-level energy, $E_u$, defines a straight line in the excitation diagram with a slope 1/T (Goldsmith & Langer 1999). Since we only have three line detections (S(0), S(1), S(2) ) the excited $H_2$ total column density, $N(H_{2tot})$ and excitation temperature, can be determined with very modest precision. Observations of higher J transitions (at shorter wavelengths) would be needed to obtain a precise determination of the excitation temperature. It is well known that the ortho-$H_2$ levels (S(1)) have systematically lower $N_u/g_u$ than the adjacent para-$H_2$ levels (S(0), S(2)) producing a "zigzag" distribution (see e.g. Fuente et al. 1999) frequently seen in the excitation diagram of photo-dissociation regions. This can also be seen in Fig. 4. Detailed physical modelling of the data in this figure is prevented by the fact that flux measurements were performed on the combined spectrum from observations of various regions in the inner part of IC 348. Each region may have a different temperature and require different extinction corrections (the latter are expected to be small).

An estimation of the excitation temperature of the gas in the inner region of IC 348 is very important in order to guide the identification of organic molecules in its diffuse gas. This was obtained from the slope of the straight line defined by the two available para-$H_2$ levels and resulted $T_{exc}$= 270 K, a fit to the three lines does not change this estimate by more than 10 K, but it would need proper modelling to account for the ortho-to-para ratio and for possible deviations from thermal equilibrium. This study is out of the scope of this paper. Adopting $T_{exc}$= 270 K, a total molecular hydrogen column density $N(H_{2tot})$ = 2.3 x $10^{21}$ cm$^{-2}$ is obtained from expression 2. Interestingly, Snow et al. (1994) obtained $N(H_2)$ = 5.5 x $10^{20}$ cm$^{-2}$ (and N(H I)= 2.6 x $10^{21}$ cm$^{-2}$ ) from observations in the line of sight of the most luminous IC 348 star which is located in the centre of the region. The consistency of these determinatinos within the uncertainties suggest that the adopted approximations provide reasonable estimates of the physical conditions of the gas.

**4.1.2 HD**

The singly deuterated molecule, HD, has a small dipole moment (approx. $10^{-4}$ Debye) and its lowest J=1-0 line at 112 $\mu$m was already detected with ISO (Wright et al. 1999). The (0-0) R(5) J=6-5 line at 19.5 $\mu$m was observed by Bertoldi et al. (1999) in the Orion shock and in combination with $H_2$ data used to determine the D/H ratio in this star-forming region resulting in [D]/[H] = 1-2 x $10^{-5}$ .

The IC 348 spectrum was searched for the presence of the series of pure rotational HD lines 0-0 R(3) to R(9) (with wavelengths 28.502, 23.034, 19.431, 16.894, 15.251, 13.593 and 12.472 $\mu$m, respectively). From their Einstein coefficients, excitation energies and the general low abundance of deuterium in the ISM all these lines were expected to be weak and they seem to be severely blended with other stronger emission features making imposible any reliable determination of fluxes.

## *4.2 Atomic* species

**Hydrogen**
The search for atomic hydrogen in the IC 348 gas results on the posible presence of the hydrogen line at 32.378 $\mu$m. This emission feature has peak strength of 0.04 Jy relative to the continuum and integrated flux of 6.6 x $10^{-18}$ W m$^{-2}$. Other hydrogen lines at 28.34 and 30.21 $\mu$m are either not detected or blended with stronger bands that make imposible to assess their presence in the observed spectrum. The hydrogen recombination lines H7$\alpha$ (HI 8-7) 19.062 $\mu$m, H6$\alpha$ 12.372 $\mu$m, H7$\beta$ 11.309 $\mu$m and H8$\gamma$ 12.385 $\mu$m are not detected. We note a strong emission line at 12.36 $\mu$m but the wavelength difference with H6$\alpha$ appears too large to support an association. The recombination line HI 13-9 at 14.183 $\mu$m could be contributing to a strong line (peak flux 0.035 Jy) seen at 14.175 $\mu$m. No evidence is found for any other lines of atomic hydrogen in the 10-30 $\mu$m spectral range.

**Sulphur**
A search for the atomic sulphur [SI] $^3P_1 \rightarrow ^3P_2$ 25.249 $\mu$m line resulted in no detectable feature with strength above 0.01 Jy. The line at [S III] 18.71 $\mu$m is possibly detected with peak flux of 0.007 Jy, however the line at 33.48 $\mu$m is not detected (peak flux below 0.006 Jy). The ratio [SIII]18.71/ [S III] $\lambda$33.48 is a density diagnostic. CLOUDY models (Ferland et al. 1998) show that a ratio > 0.6 corresponds to a density of > 100 cm$^{-3}$.

**Iron**

The well known [Fe II] 25.988 μm line could be responsible for an emission feature at 25.99 μm (peak flux of 0.03 Jy). However, the 24.519 μm and the 17.94 μm [Fe II] lines are not detected in the spectrum.

**Neon**

The [Ne II] 12.81 μm line is possibly present in the spectrum but it coincides with an OH and $C_2H_2$ 12.806 μm band. Given the spectral resolution of our spectrum it is not possible to separate the relative contributions of these bands. The ionization potential of this Ne emission line is 21 eV and thus it would indicate that the slit locations may have captured emission from an HII region. The [Ne III] line at 15.56 μm is also seen in emission with a flux peak of order 0.03 Jy (excitation temperature of the upper level is approx. 940 K). The [Ne V] line at 14.32 μm is however not present.

**Oxygen.**

The [O IV] line at 25.89 μm could be present in the IC 348 spectrum albeit with a peak flux below 0.007 Jy, only slightly above the detection limit.

In summary, the probable presence of weak lines of ionized atomic species of sulphur, iron and neon suggests that in some of the observed zones of the interstellar gas which contribute to the spectrum there is a component of gas at relatively high excitation temperature.

## 4.3 Diatomic molecules

Absorption spectra of diatomic molecules commonly reported in the literature were computed using the python package HAPI and HITRAN. Fig. 5a and 5b show the synthetic spectra for a temperature of 240 K, adopted as representative of the gas in IC 348. A search for the strongest bands of each molecule was performed in the observed IC 348 spectrum. If the result was positive, then a more detailed analysis of weaker bands of that molecular specie was carried out.

**4.3.1 OH Hydroxile**

The relative strength of the OH lines in the 10-30 $\mu$m spectral region is quite sensitive to temperature according to HITRAN computations. For instance, below 250 K the strongest theoretical band appears at 28.94 $\mu$m, and also there are relatively strong doublets at 30.28, 30.35 and 30.66, 30.70 $\mu$m. None of these bands are clearly detected in the spectrum of IC 348, nor are the weaker bands predicted at shorter wavelengths. In Fig. 6a, it is plotted a spectral synthesis for OH assuming a temperature of 300 K. The molecular hydrogen analysis indicated that there is gas at this temperature contributing to our IC 348 spectrum and indeed the predicted OH transitions at 300 K are detected in the observed spectrum with relative intensities consistent with the predictions.

We also detect a sequence of highly excited OH (v = 0, J→J−1) pure rotational transitions arising in the $^2\Pi_{3/2} \rightarrow {}^2\Pi_{1/2}$ rotational ladders (see wavelengths in Table 3 of Najita et al. 2010) which originate from high energy levels and have large transition probabilities (A-values of 20-70 s$^{-1}$). Some of the OH lines more clearly seen in our spectra, depicted in Figures 6a and 6b, correspond to high J values in the range J=29.5 to J=20.5 (with upper energy levels up to 4,000 K). Lines with lower J values appear weak or blended with other emission features in the spectrum.

It is unlikely that the chemical formation route via $H_2 + O \rightarrow OH + H$ (e.g. Hollenbach & McKee 1979; Neufeld & Dalgarno 1989) is the dominant source of OH. This reaction has Arrhenius activation energies $E_A/k_B$ about 3150 K and only proceed efficiently in dense, hot gas. UV induced photodesorption of water ice from grain mantles and photodissociation of $H_2O$ either in the gas phase or directly in the grain ice mantles (see Andersson et al. 2006, Tappe et al. 2008) could be the primary sources of OH in the IC 348 gas.

**4.3.2 CO**

Carbon monoxide (CO) is the second most abundant molecule in the gas phase of the interstellar medium. In dense molecular clouds, it is also present in the solid phase as a constituent of the mixed water-dominated ices covering dust grains. Some rotational lines appear in our HITRAN based computed absorption spectrum, the most prominent are R (75), R (76) and R (77) at 35.51, 35.09 and 34.66 µm, respectively. These bands could be contributing to emission features in the IC 348 gas spectrum, but they are severely blended and it is not possible to ascertain their relative contribution against other contaminating bands. No other CO bands seem to be present in the IC 348 spectrum.

**4.3.3 $N_2$**

Isolated $N_2$ molecules are IR-inactive, and solid $N_2$ has only weak IR activity, hindering its astronomical detection by IR methods. The discovery of a new $N_2$ IR feature would be of great interest, the HITRAN synthetic spectrum shows that the strongest features appear at 11-11.5, 13.11, 15.46, 17.23, 21.84 and 22.83 µm. None of these bands could be unambiguously detected in our spectrum. The series of transitions between 11 and 11.5 µm are masked by the strong PAH 11.2 µm emission. The 13.11 µm band is barely resolved from the OH line at 13.08 µm, the 15.46 µm is blended and very weak with peak flux below 0.006 Jy, thus unreliable, the 17.23 µm line, if present, would be a small contribution blended with a relatively strong $H_2O$ 17.225 µm water line (expected from the computations for water, see below).

*4.4 Tri-atomic molecules*

Figure 7 shows HITRAN based computations in the range 10-19.5 µm for some of the most relevant triatomic molecules detected in diverse astrophysical environments: $H_2O$, $CO_2$, HCN, $NO_2$, $N_2O$. In the spectral range 20-32 µm only water presents significant bands.

**4.4.1 $H_2O$**

Water has dipole-allowed pure rotational transitions that occur at mid/far-infrared and submillimeter wavelengths which have been extensively reported (Liseau et al. 1996, Harwit et al. 1998, Wright et al. 2000). In star-forming regions water is ubiquituous (e.g. van Dishoeck et al. 2021) with variations in its abundance relative to $H_2$ ranging from $< 10^{-8}$ in the coldest clouds up to values $> 10^{-4}$ in warm gas and shocks. The HITRAN computation for a temperature of 250 K shows a large number of water transitions in our spectral range. In Fig. 8a and 8b, the dependence with temperature is displayed. Many of these transitions have been previously observed in astrophysical environments (e.g. Pontoppidan et al. 2010, Blevins et al. 2016) and are also detected in our IC 348 gas spectrum with fluxes listed in Table 4. Fig. 9a shows water synthetic bands as compared with the observed spectrum in the range 12-13 µm. No evidence for significant water contamination is found in this spectral range. Water bands at 28.591, 22.639, 22.538, 17.358, 17.225 and 17.103 µm are detected (see Fig. 9b). The bands at 30.525 and 30.529 µm are however not deteceted. At the moderate resolution of our spectrum, some of the water line complexes are severely blended rendering it impossible to measure fluxes for individual transitions. Where it was possible, band fluxes were calculated by defining and subtracting a linear fit to the local continuum and fitting Gaussians to the lines. Lines with peak fluxes of less than 0.007 Jy are considered below the detection limit.

The water rotational diagram (Fig.9c) was produced with the observed fluxes and the molecular parameters listed in Table 4. Only lines detected with reliable flux measurements were used. The lines at 12.444 and 12.453 µm are severely blended at the spectral resolution of IRS, it is not possible to disentangle which fraction of the flux ascribed in Table 4 to the 12.444 line µm is contributed by the 12.453 µm and therefore they do not appear in the diagram even though a water feature is clearly present in the spectrum. From the slope of the least square fit plotted in the figure we obtain an excitation temperature of T=385 K and from the value of the ordinate at the origin and using the corresponding partition function value given by HITRAN, we derive a total water column density of $N_{tot}$ ($H_2O$) = 2.7 $\times 10^{15}$ cm$^{-2}$. The resulting ratio [N($H_2O$)]/[N($H_2$)] ~ $10^{-6}$ is not unusual in regions with warm gas and shocks (van Dishoeck et al. 2021).

While there is a significant presence of water bands in our spectrum, the many high excitation temperature transitions > 4000 K of water known to exist in our spectral range are too weak for any detection given the resolution and sensitivity of the present spectrum.

### 4.4.2 $CO_2$

The strongest $CO_2$ band in our spectral range as can be seen in Fig. 10 is the well known $v_2$ 1–0 14.98 µm band. Detections of this band are extensively reported in the literature (van Dishoeck et al. 1996, van Dishoeck 1998, Dartois et al. 1998, Boonman et al. 2003) in a variety of star-forming regions. The $CO_2$ abundances with respect to $H_2$ obtained are typically of order 10-7, up to two orders of magnitude lower than those of solid $CO_2$ in the same regions.

The 14.98 µm band is clearly seen in the IC 348 spectrum with a peak flux of 0.017 Jy with respect the local continuum. The other $CO_2$ bands in the computed spectrum are significantly weaker with expected peak fluxes below 0.01 Jy. Some of these bands could indeed be present, e.g. the doublet at 13.876, 13.898 µm at peak flux about 0.007 Jy, but close to the detection limit. Many other weak $CO_2$ lines predicted by the HITRAN computation, are probably contributing to the confusion level at fluxes of order 0.005 Jy. These individual lines cannot be resolved or identified at the resolution and sensitivity of our spectrum. We measured a flux of $9 \times 10^{-18}$ W m$^{-2}$ for the strongest line at 14.98 µm. If we assume a gas temperature of 270 K, the column density of $CO_2$ has a value of $5 \times 10^{12}$ cm$^{-2}$. The other weaker lines of $CO_2$ that we could tentatively identify at 13.88 and 13.90 µm with lower fluxes make a factor 2 smaller contribution to the total column density of $CO_2$ which we estimate is of order $4 \times 10^{13}$ cm$^{-2}$.

### 4.4.3 HCN Hydrogen cyanide

The HITRAN computation displays a rich HCN spectrum in the wavelength range of our interest (see Fig.11) with the strongest band being the $v_2$ 1-0 fundamental ro-vibrational band at 14.045 µm (see e.g. Lahuis and van Dishoeck 2000). We find in the IC 348 spectrum a relatively strong feature (0.01 Jy peak flux) at 14.04 µm which could be associated to this HCN band (blend of transitions 12_76 → 11_47 and 10_73 →9_46). The next most intense HCN band according to the computation may be contributing to the emission feature seen at 13.977 μm. However, given the relative strength with respect to the 14.045 μm band, its contribution to this feature should be minor. The weaker HCN feature at 13.76 μm seems to be present and well resolved in the IC 348 spectrum with a rather low peak flux of 0.007 Jy. Other HCN bands may also be present at similar strength, i.e close to the detection limit. At such small flux level all of them appear blended with other emission features.

Hydrogen isocyanide HNC is expected to be less abundant than HCN, the strongest HNC $v_2$ band at 21.54 μm is not detected in the IC 348 spectrum.

### 4. 4.4 $NO_2$

The HITRAN computation shows that the strongest band of this molecule is located at 12.65 μm. At this wavelength, there is a very weak feature at noise level in the IC 348 spectrum. Since the other bands in the synthesis are considerably weaker, no contribution is expected from this molecule to any emission lines above the detection limit.

### 4.4.5 $N_2O$

The HITRAN computation shows the strongest band of this molecule is located at 16.97 μm. This band is below the detection level in the IC 348 spectrum. The remaining synthetic bands are considerably weaker, thus no significant contribution to any emission feature is expected from this molecule.

## *4.5 Tetra-atomic molecules*

Fig. 12 shows the synthetic spectra obtained with HITRAN for $C_2H_2$, $NH_3$ and $SO_3$.

### 4.5.1 $C_2H_2$  Acetylene

$C_2H_2$, a well-known precursor of PAHs and other symmetric hydrocarbons is quite abundant in the interstellar medium. These symmetric hydrocarbons do have permitted infrared vibration-rotation transitions. The $v_5$ band has three branches, P(J-1), Q(J), R(J+1) where J is the rotational quantum number of the upper vibrational state. The $v_5$ band at 13.71 μm is the strongest ro-vibrational band of $C_2H_2$ (symmetric bending mode in which the two H atoms vibrate together) detected in diverse astrophysical environments (Lacy et al. 1998, Lahuis and van Dishoeck 2000, Boonman et al. 2003, Carr and Najita 2008, Najita et al. 2021). A well resolved emission feature in the IC 348 spectrum with a peak flux of 0.008 Jy with respect to the local continuum is found at 13.71 μm (see Fig. 13) and could be caused by this $C_2H_2$ band. The measured flux of this band is $2 \times 10^{-18}$ W m$^{-2}$. If we assume this molecule is at the same temperature as the $H_2$ it would lead to a column density of $N(C_2H_2)$ of order $1 \times 10^{12}$ cm$^{-2}$. The R(23) $C_2H_2$ at 12.73 μm could also be present, blended with the R(24) HCN line. Other $C_2H_2$ bands in the HITRAN-based computation are predicted to be much less intense than these two bands, thus below the detection limit of our spectrum.

### 4.5.2 NH$_3$

NH$_3$ bands reported in the mid-IR (see e.g Najita et al. 2021) were searched in the IC 348 spectrum. The 10.74-10.76 μm emission feature (peak flux 0.01 Jy, see Fig. 14) can be associated to NH$_3$ antisymmetric transitions aQ(3,3), aQ(4,4), aQ(7,6). In addition, evidence for NH$_3$ bands at 10.50 and 10.35 μm is also found as weak emission features (with peak flux 0.007 Jy and 0.005 Jy, respectively). From these emission lines we derive $N(NH_3) = 1.3 \times 10^{14}$ cm$^{-2}$. For comparison, using the Green Bank radiotelescope Rosolowsky et al. (2007) reported as typical column density of $N(NH_3) \approx 10^{14.5}$ cm$^{-2}$ for the cores of the Perseus Molecular Cloud.

### 4.5.3 CH$_3$

The methyl radical CH$_3$ is an important intermediate product in the basic ion-molecule gas-phase chemistry networks in the interstellar medium driven by cosmic-ray ionization (Herbst & Klemperer 1973) and it is an important building block in the formation of small hydrocarbon molecules. Together with CH and $CH_2$, it is produced by a series of reactions starting with $C + H_3^+ \rightarrow CH^+ + H$ or via the radiative association of $C^+ + H_2 \rightarrow CH_2 + h\nu$, followed by hydrogen abstraction reactions and dissociative recombination. Alternatively, $CH_3$ can be produced by the photodissociation of methane ($CH_4$). Subsequent reactions of $C^+$ with $CH_3$ form one of the most important steps in the formation of small hydrocarbons.

It is interesting to search for CH, $CH_4$ and $C_2H_2$ along the same line of sight as it traces basic gas-phase astrochemistry networks. CH was reported by Snow et al. (1994) in the line of sight towards the centre of IC 348 N(CH)=5.3 x $10^{13}$ cm$^{-2}$ and $C_2H_2$ is also present as discussed above.

The strongest $CH_3$ Q–branch transition is located at 16.5 µm and another strong transition is the R(0) line at 16.0 µm. There is a clear blended emission feature in the IC 348 spectrum at 16.00 µm and at 16.49 µm we find another strong-blended emission feature (flux ~1.6 $10^{-17}$ Wm$^{-2}$ ). Both emissions could have contribution from $CH_3$. The weaker R(1) at 15.54 µm could be responsible for the blue asymmetry in the strong emission feature seen at 15.56 µm which, as previously discussed, is dominated by a combination of [Ne III] and H2O . The $CH_3$ P(2) at 17.60 µ is, unfortunately weak and below the detection limit. In summary, while it is possible that the strongest transitions of $CH_3$ contribute to emission features in the IC 348 spectrum, the existing blends preclude an estimation of the abundance of this molecule. Gas-phase $CH_3$ has been detected through its $\nu_2$ bending mode lines at 16.0 and 16.5 $\mu$m along the line of sight toward Sgr A* with an abundance of $10^{-8}$ (Feuchtgruber et al. 2000).

*4.6 Polyatomic Molecules*

In Fig. 15 we plot the HITRAN synthetic spectra for a number of molecules with five atoms or more.

**4.6.1 Five atoms**

**CH₄ Methane**

Lacy et al. (1991) detected the CH$_4$ R(0) and R(2) lines with ground based observations toward NGC 7538 and Dartois et al. (1998) and (Boogert et al. 1998) reported observations of gaseous and solid phase CH$_4$ toward the embedded protostar GL 7009S. These are bands at wavelengths shorter than those covered by our spectrum. There are no relevant bands of CH$_4$ in the IC 348 spectrum which we may expect to detect.

**HC₃N Cyanoacetylene**

This molecule is one of the building blocks from which more complex hydocarbons are produced (Cernicharo et al. 2001). According to the HITRAN based synthetic spectrum in Fig. 15 and to Yamada & Bürger (1986), the strongest band of HC$_3$N is $v_5$ located at 15.10 µm. This band could be responsible of a weak emission feature found at 15.11 µm in the IC 348 spectrum with peak flux of 0.007 Jy, very close to the detection limit of our spectrum (see Fig. 16).

**4.6.2 Six atoms**

**C₂H₄ Ethylene**

There is a clear emission feature at 10.53 µm (peak flux 0.011 Jy) which may be caused by the 10.53 µm Q branch of the $v_7$ CH$_2$ wagging-mode of the C$_2$H$_4$ (Cernicharo et al. 2001, Bast et al. 2013). The strongest line in the HITRAN computation is located at 15.9 µm (see Fig. 15), as we will see below this is also closely located to the strongest line of C$_4$H$_2$, and both lines could be contributing to the broad but weak emission feature at 15.84-15.90 µm seen in the IC 348 spectrum with a peak flux of order 0.01 Jy.

**C₄H₂ Diacetylene**

Cernicharo et al. (2001) reported the detection of C$_4$H$_2$ in ISO observations of CRL 618. The absorption signal near 15.9 µm was identified and assigned to the $v_8$ 15.87 µm fundamental bending mode of $C_4H_2$ based on the laboratory work of Arié & Johns (1992). In the IC 348 spectrum there is an emission feature at 15.87 µm with peak flux of 0.01 Jy, but it is severely blended with other emission bands of similar strength (see Fig. 16 ).

### 4.6.3 Seven atoms

**$CH_3C_2H$ propyne or methylacetylene**

It has its strongest band at 15.79 µm. No emission feature could be found at this wavelength.

**$HC_5N$**
It is expected that $HC_5N$ (like $HC_3N$) is formed by the reaction of $C_4H_2$ + CN. As we may have found evidence for $C_4H_2$ it is appropriate to search for $HC_5N$. According to Deguchi and Uyumura (1984), the strongest transition of this molecule in our spectral range (see Fig. 15) is $v_7$ at 14.60 µm (685 cm$^{-1}$), followed in strength by the $v_8$ 17.67 µm (566 cm$^{-1}$). Both lines seem to be present in the spectrum of the interstellar gas of IC 348, albeit rather weak (peak flux of 0007 Jy) and possibly blended with other similarly weak lines (Fig. 16).

### 4.6.4 Eight atoms

**$C_2H_6$ Ethane**

$C_2H_6$ is an important molecule in planetary and cometary atmospheres (see e.g. Burgdorf et al. 2006, Coustenis et al. 2003). It has strong rotational-vibrational bands in the infrared, and the $v_9$ band at 12 µm (Fig.15) has been widely studied (Daunt et al. 1981, Vander Auwera et al. 2006).
The HITRAN computation shows a broad system of lines associated to the $v_9$ mode with the strongest transition at 12.16 µm (822.35 cm$^{-1}$). A relatively broad feature (0.03 µm) with peak intensity 0.009 Jy is detected at 12.15 µm in the IC 348 spectrum which could be due to this $v_9$ band of $C_2H_6$.

**$C_6H_2$ Triacetylene**

According to laboratory work ( e.g. Haas et al. 1994) the $v_{11}$ fundamental bending mode of $C_6H_2$ is produced at 16.10 µm. Cernicharo et al. (2001) reported the detection of this band in ISO observations of CRL 618. In the IC 348 gas spectrum there is a clear well resolved emission feature at 16.10 µm which we abscribe to $C_6H_2$ (see Fig. 16).

**4.6.5 More than 9 atoms.**

**$C_6H_6$ Benzene**

The bending mode $v_4$ (all hydrogens moving away in the same direction from the plane formed by the six carbons) is the strongest infrared band of $C_6H_6$. The lines in the $v_4$ Q-branch are overlapped, producing a relatively strong narrow feature at 14.837 µm (Lindemayer 1988, Cernicharo et al. 2001). In addition to the $v_4$ transition, there is a hot band at 14.862 µm, $v_4 + v_{20} \leftarrow v_{20}$ where $v_{20}$ is the lowest frequency vibrational mode of benzene at 398 cm$^{-1}$. The vibrational band strength for the hot band could be similar to that of the fundamental $v_4$ (see Kauppinen, Jensen, & Brodersen 1980). In the IC 348 spectrum there is a clear emission feature at 14.84 µm (with peak flux of order 0.013 Jy) which may have significant contributions from the $v_4$ and hot $C_6H_6$ bands and also from the P (13) line of HCN at 14.847 µm. The measured flux of the 14.84 µm emission feature (from the gaussian fit in Fig. 17) mainly attributable to benzene is 4 ($\pm$0.5) x 10$^{-18}$ W m$^{-2}$.

*4.7 PAHs*

The C-C and C-H stretching and bending modes of polycyclic aromatic hydrocarbons (PAHs) at 6.2, 7.7, 8.6, 11.2, 14.2, 16.4 $\mu$m dominate the mid-IR spectra of many objects and are indicators of the presence of complex carbonaceous material excited by UV radiation (Leger & Puget 1984, Allamandola et al. 1985). The PAH transitions in our range of interest: 11.2, 12.0, 12.7, 14.2, and 16.4 µm correspond to C-C modes. The strongest PAH band observed in our spectrum is at 11.2 µm.

*4.8 Fullerenes*

The detection and determination of fullerene abundances in various interstellar regions of IC 348 was reported by Iglesias-Groth (2019) using Spitzer IRS data. Several bands of $C_{60}$, $C_{70}$ and $C^+_{60}$ were identified. The spectrum under study here involves various other regions of the interstellar medium in IC 348 and several emission features are also found consistent with known fullerene bands (peak fluxes are given in parenthesis), most remarkably: for $C_{60}$ 17.33 $\mu$m (0.02 Jy) and 18.88 $\mu$m (0.06 Jy); for $C_{70}$ 12.63 $\mu$m (0.01 Jy), 13.83 $\mu$m (blended), 14.90 $\mu$m (0.012 Jy), 15.63 $\mu$m (0.03 Jy), 17.78 $\mu$m (0.035 Jy) and, for $C^+_{60}$ 10.46 (0.015 Jy), 13.22 (0.012 Jy) and 18.58 $\mu$m (0.03 Jy).

*4.9 Ices*

Water was the first molecule detected in the solid state in the interstellar medium (Gillet and Forrest 1973). Since then many other molecules have been identified in icy form (i.e. CO, $CO_2$, $CH_4$, $NH_3$ and $CH_3OH$). These ices are known to be common during the cold and dense stages of star formation, with abundances reaching $10^{-4}$ n($H_2$). Icy dust grains play a key role in the formation of more complex organic molecules (COMs), such as glycolaldehyde ($HOCH_2CHO$) and ethylene glycol ($HOCH_2CH_2OH$). While there is clear evidence that the combined ISM IC 348 is dominated by emitting warm gas, the possibility of ices in some of the regions used in the construction of this spectrum should not be discarded. In the following paragraphs, we report a search for bands from the most common ices.

**$H_2O$.**
The most direct evidence for solid-state formation of interstellar water comes from the detection of the 3 $\mu$m O-H stretching vibration band of water ice toward numerous infrared sources. In many cases, the 6 $\mu$m and 11 $\mu$m librational modes have also been observed. Giuliano et al. (2014) presents laboratory spectra of amorphous and cristallyne water in the spectral range of our interest, both spectra display 10-20 $\mu$m wide bands which are not seen in our spectrum of the gas in IC 348.

**CO₂**

Based on laboratory spectroscopy of mixed $CO_2$ ices (D'Hendecourt & Allamandola 1986) the $\nu_2$ bending mode of $CO_2$ ice in absorption at 15.20 μm was identified by D´Hendecourt and Jourdain de Muizon (1989) toward several sources in the IRAS database. Ehrenfreund et al. (1997) showed the existence of new bands associated to this bending mode when $CO_2$ is present in a mixture of other ices. Multiple laboratory experiments measured the shifts and the presence of various bands as a function of the mixture. In several cases three bands of similar strength were found around 15.2 μm. In the IC 348 gas spectrum we find three strong well resolved emission features in this region at 15.09, 15.14 and 15.21 μm ( 662.7, 660.5 and 657.5 $cm^{-1}$, respectively) which fit well the $CO_2$ ice bands in the case of a mixture of CO:O2:CO2 (see Table 3 in Ehrenfreund et al 1997). There may be other mixtures that could also reproduce these features, so this is only a possible explanation for the bands detected between 15.0 and 15.2 μm.

**NH₃**

Giuliano et al. (2014) give measurements of $NH_3$ ices and other ice transitions in our spectral region. For $NH_3$, the reported bands are at 439 $cm^{-1}$ (23.87 μm) for amorphous ice, and 535 and 421 $cm^{-1}$ for crystalline ice (18.69 μm and 23.75 μm, respectively). A weak band in the IC 348 spectrum at 23.87 μm could be due to amorphous $NH_3$ ice. The crystalline bands are undetected or severely blended with bands of other species.

**CH₃OH**

$CH_3OH$ and other ices are proposed sources of complex organic molecules (Charnley et al. 1992; Garrod et al. 2008). The determination of ice abundances and production channels in star-forming regions is very relevant for studies of prebiotic chemistry. Hudgins et al (1993) measured the $\nu_{12}$ mode of $CH_3OH$ at 705 $cm^{-1}$ (14.184 μm). An emission feature at 14.175 μm in the spectrum of IC 348 could be associated to this band. The amorphous $CH_3OH$ 330 $cm^{-1}$ (30.30 μm) band could be associated to a rather strong feature in the IC 348 spectrum with peak flux 0.04 Jy. The crystalline 347 $cm^{-1}$ band at 28.82 μm is possibly detected, although with a modest peak flux of 0.008 Jy.

## 5. DISCUSSION and CONCLUSIONS

The mid-IR spectrum resulting from observations conducted by Spitzer at several interstellar locations in the core of the IC 348 young star cluster revealed emission bands from a large variety of molecular species. Each observed location is subject to different UV radiation fields and very probably the combined averaged spectrum shows emission lines of molecules under a variety of physical conditions. While averaging spectra from different regions limits the possibility of a detailed chemical composition analysis, the much higher S/N of the combined spectrum opens an interesting opportunity to identify weak bands from relevant molecules for the chemistry of the interstellar medium which would not be revealed by individual spectra.

The averaged IC 348 spectrum shows emission lines that can be confidently associated to the vibration-rotation bands of very common molecules in the ISM: $H_2$, $H_2O$, $CO_2$ and $NH_3$. The identification of rotational $H_2$ lines allowed us to derive a column density for molecular hydrogen $N(H_2) \sim 2.3 \times 10^{21} cm^{-2}$ which is consistent with previous determinations of hydrogen content in the core of IC 348 by Snow et al. (1994). The derived excitation temperature $T_{ex} \sim 270 \pm 30$ K for molecular hydrogen in IC 348 is an indication of warm gas in the inner region of the cluster, possibly associated to some photon-dominated regions. A few spectra were taken in locations close to the B-type star LRLL 2 and its UV radiation could be the main cause for the excitation of the gas. The temperature excitation diagram of $H_2O$ suggests the existence of even warmer regions with temperatures of order 385 K. From the observed water lines, we derived a column density $N(H_2O) \sim 3 \times 10^{15}$ cm$^{-2}$, and therefore, an abundance relative to molecular hydrogen $[H_2O]/[H_2] \sim 10^{-6}$. For reference, this ratio is similar to the lowest values found for the inner hot cores of high mass protostars and a factor 10 higher than the values measured in their cold envelopes (van Dishoeck et al. 2021).

The detection of transitions of weak highly excited OH also suggests the presence of regions with hot and dense gas able to populate the upper levels of this molecule via collisions. This highly excited OH emission may result from the photodissociation of $H_2O$ by the stellar UV radiation (see e.g. Mandell et al. 2008).

$CO_2$ is predicted to be among the more abundant carbon- and oxygen-bearing gas-phase species in massive star-forming regions (e.g. Charnley 1997). We derive a column density of $N(CO_2)=1 \times 10^{13}$ cm$^{-2}$ in the gas of IC 348 and a ratio $[CO_2]/[H_2]$ ~$10^{-8}$. In Van Dishoeck et al. (1996) gas-phase $CO_2$ abundances relative to $H_2$ of order $10^{-7}$ were commonly reported and abundances a few times $10^{-7}$ have also been found in the direction of Orion-IRc2/BN, however an order of magnitude lower abundances were also found toward the shocked regions Peak 1 and 2 (Boonman et al. 2003). At T>230–300 K most of the OH is probably driven into $H_2O$, thus reducing the formation of $CO_2$ through its primary formation route in gas-phase via the reaction CO + OH →$CO_2$ + H. This could be an explanation for the relatively low abundances of $CO_2$ in the gas of IC 348.

$CO_2$ is also commonly detected in interstellar ices, with abundances of 15% with respect to $H_2O$ ice, or ~ $10^{-5} - 10^{-6}$ with respect to $H_2$ (e.g. Gerakines et al. 1999). As we find emission lines between 15.0 and 15.2 $\mu$m which are consistent with the laboratory bands measured by Ehrenfreud et al. (1997) in the case of an ice mixture of CO:O2:CO2 it is possible that some of the observed regions in IC 348 contain cold enough gas to maintain a significant abundance of $CO_2$ in ice form.

Observations of $C_2H_2$ (acetylene) and HCN (hydrogen cyanide) are interesting because they are both relevant in the carbon- and nitrogen chemistry, and because their excitation provides information on the physical conditions of the medium (Lahuis & van Dishoeck 2000). Acetylene, a key ingredient in the production of large complex hydrocarbon molecules (Herbst 1995) has been observed toward low- to high-mass star-forming regions, and detected in the gas phase either in absorption (e.g., Carr et al. 1995; Lahuis & van Dishoeck 2000) or in emission (e.g., Boonman et al. 2003) mostly toward young stellar objects with excitation temperatures ranging from 10 to 900 K and abundances with respect to $H_2$ ranging from a few times $10^{-8}$ to a few times $10^{-7}$.

Assuming an excitation temperature of order 300 K for the gas in IC 348, the observed flux of $2 \times 10^{-18}$ W m$^{-2}$ for the emission feature at 13.72 $\mu$m attributed to the strongest line of the $\nu_5$ C$_2$H$_2$ 1-0 band (Einstein coefficient A=3.06 s$^{-1}$) leads to a column density for acetylene of order $10^{12}$ cm$^{-2}$ and abundance relative to molecular hydrogen of $10^{-9}$. This abundance is very close to values reported by Sonnentrucker et al. (2007) toward various lines of sight in the star-forming region Cepheus A east, where C$_2$H$_2$ emission mainly arises in a warm component mostly located in front of the cold quiescent gas. These authors find a good correlation between the column densities of C$_2$H$_2$ and CO$_2$ that extends over two orders of magnitude in column density. The gaseous [C$_2$H$_2$]/[CO$_2$] ratios they derived for the different lines of sight were in the range 0.04-0.2 and the mean value was 0.08. For the gas in IC 348 we derive a ratio [C$_2$H$_2$]/[CO$_2$]= 0.03 consistent within errors with the findings in Cepheus A. Reactions of C$_2$H$_2$ with C$^+$, C and small radicals lead to long unsaturated carbon chains, whereas reactions with CN produce cyanopolyynes such as HC$_3$N (Herbst 1995) for which we only find a very marginal evidence in IC 348.

HCN is one of the more abundant nitrogen–bearing molecules in dense clouds. The $\nu_2$ HCN bending mode at 14.0 $\mu$m has been observed from the ground at high spectral resolution see e.g. Lacy et al. (1989), Evans et al. (1991) and Carr et al. (1995), and at a more modest resolution space observations revealed abundances relative to molecular hydrogen in the range $0.2\text{-}4 \times 10^{-7}$ in YSOs (Lahuis and van Dishoeck 2000). HCN has been detected in the inner few AU of disks (Carr & Najita 2008, Lahuis et al. 2006). The inferred abundances of $10^{-5}$–$10^{-6}$ are orders of magnitude larger than those found in the outer disk or in the cold surrounding cloud. In the gas of IC 348, assuming the same excitation temperature as H2 we obtain an abundance of $10^{-7}$ very similar to those reported for YSOs. These low abundances with respect to findings in protoplanetary disks are probably the result of an active participation of HCN in the high–temperature gas–phase chemistry. This is a key molecule in the formation of prebiotic building blocks, including amino acids and nucleobases.

The $C_2H_2$ and HCN abundances make it plausible that $CH_3$ be also present in IC 348 and as discussed in the previous section several emission features may be produced by this molecule, although blends prevent an abundance estimate. It is important to obtain higher resolution spectroscopy to ascertain the presence of $CH_3$. Evidence for other small molecular species such as $C_4H_2$, $C_6H_2$, $HC_5N$, $C_6H_6$ known to play a role as precursors of other more complex organic molecules suggest that the conditions in the gas of IC 348 were suitable for the activation of a rich and diverse chemistry which may have led to the production of such hydrocarbons.

Chemical reaction networks in the gas of IC 348 may have led to rather complex carbon molecules such as PAHs and fullerenes (Iglesias-Groth 2019). The strong PAH feature at 11.2 $\mu$m and the moderately strong one at 12 $\mu$m indicate the presence of large grains (Peeters et al. 1017). The comparison of our IC 348 spectrum with the IRS spectrum of the Red Rectangle (see figure 8 in Bauschlicher et al 2008) shows a great similarity in the features associated with PAHs. The bands at 11.2, 12.8, and 16.4 $\mu$m indicate the presence of long, mostly neutral PAHs, and possibly to a lesser extent of small polycyclic aromatic hydrocarbons and VSGs (Peeters et al. 2017). Abundances of fullerenes in the IC 348 gas (Iglesias-Groth 2019) suggest the existence of efficient mechanisms leading to complex carbon molecules. Murga et al. (2022) propose a path for conversion of PAHs into fullerenes in regions with an intense UV radiation field. Studies with high spatial resolution in IC 348 may show evidence for such mechanism in the proximity of the hottest stars in the star-forming regions which are the dominant sources of UV radiation.

IC 348 offers a particularly interesting chemical environment that deserves further studies at higher spectral resolution and higher sensitivity in the mid-IR. Spectroscopy with JWST may reveal the spatial distribution of the molecules for which we have reported evidence here and confirm their abundances. It is also important to extend the present search to other key molecules involved in the chemical path towards complex organic molecules which are essential for the development of life. The study of the molecules in the interstellar gas of IC 348 shall be complemented with similar studies in the protoplanetary disks of this exceptional star-forming region.

DATA AVAILABILITY

The data underlying this article were derived from sources in the public domain and are available in CASSIS, the Combined Atlas of Sources with Spitzer IRS Spectra, at https://cassis.sirtf.com/.

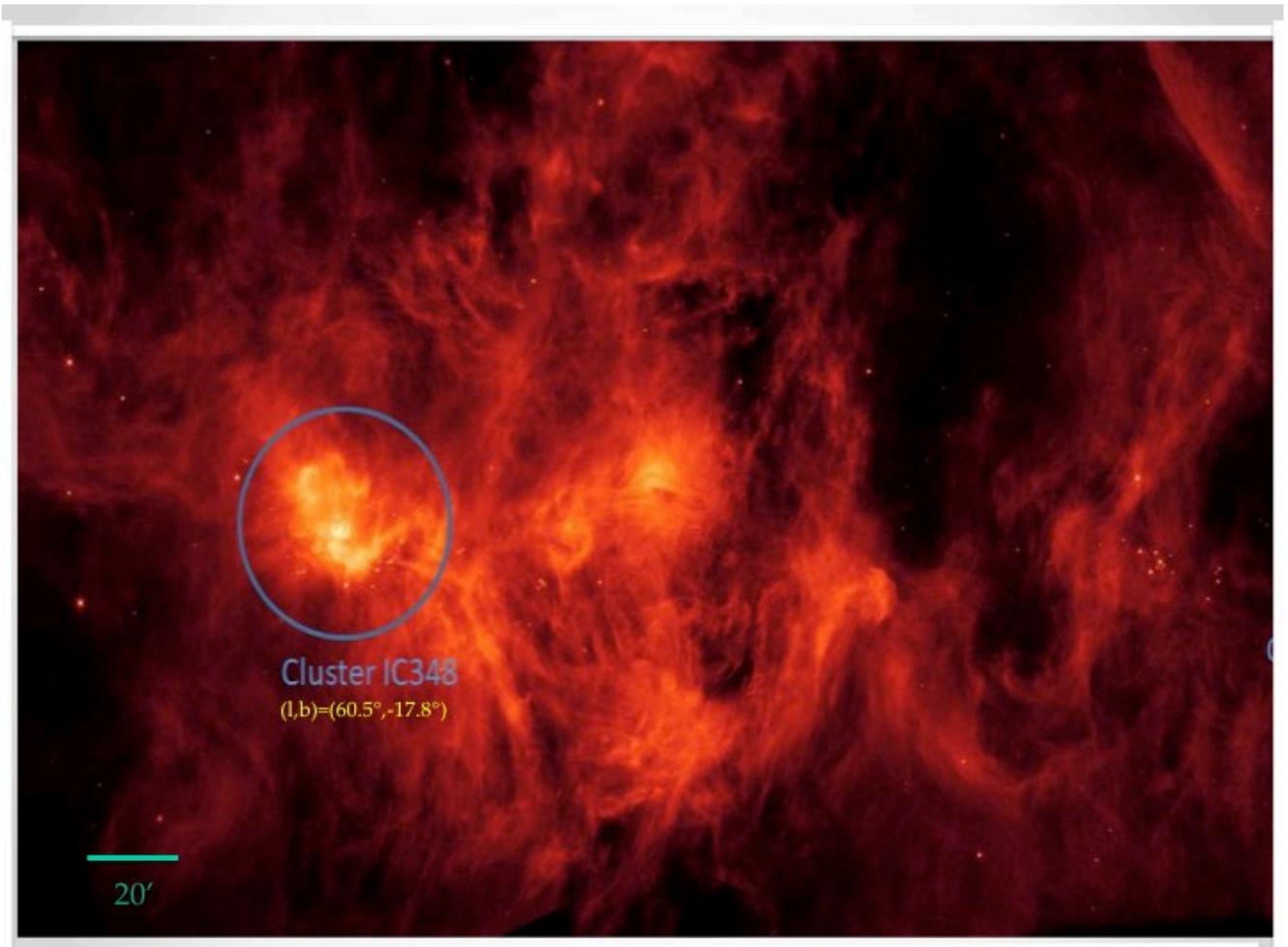

**Figure 1.**
The Perseus Molecular Cloud as observed by NASA´s Spitzer Space Telescope showing the location of the IC 348 star cluster and encircled the region containing the IC 348 pointings listed in Table 1. NASA/JPL-Caltech - https://photojournal.jpl.nasa.gov/figures/PIA23405_fig2.jpg

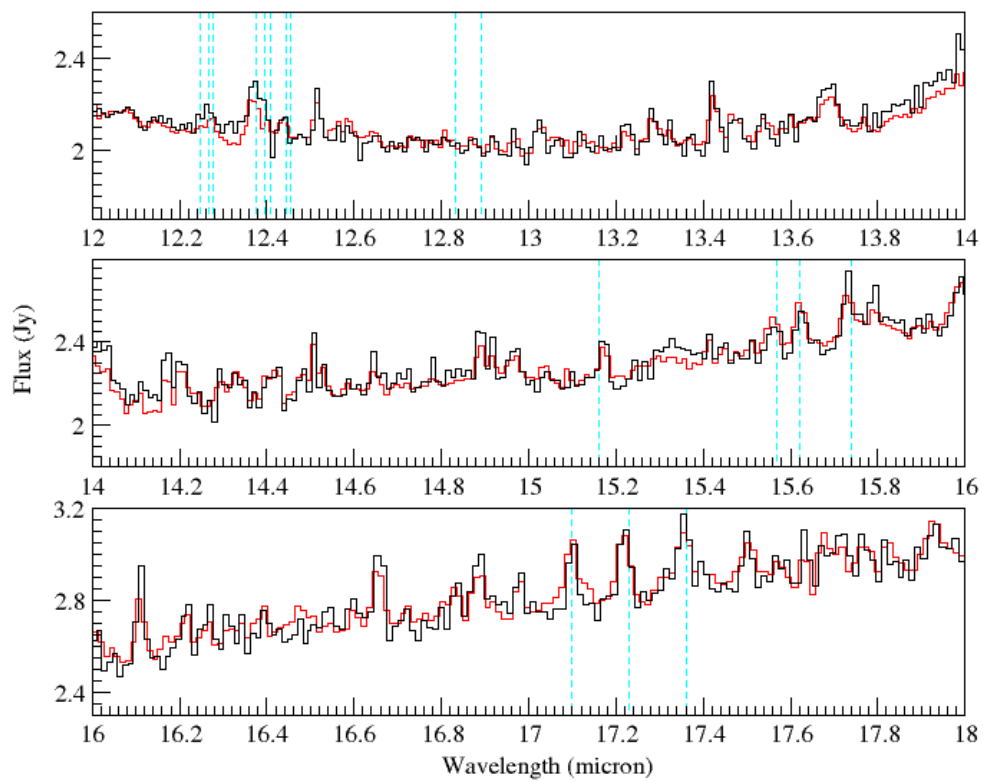

**Figure 2a.** CASSIS spectra of RNO 90 provided by the full aperture and optimal differential extraction techniques. The vertical lines indicate the location of observed $H_2O$ lines already reported in the literature for this and other sources (see Blevins et al. 2016).

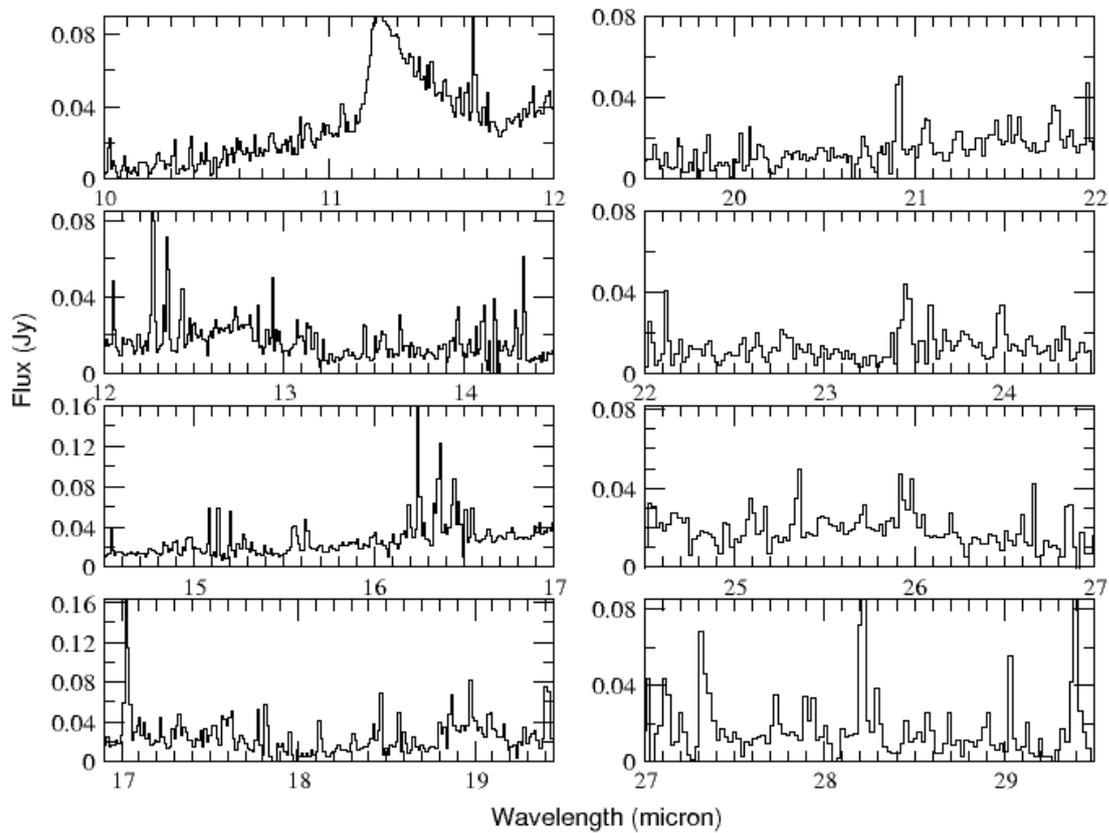

**Figure 2b**. The ISM spectrum of IC 348 in the 10-29.5 μm resulting from the combination of several independent Spitzer IRS SH (left panel) and LH (right panel) observations of the gas in various locations of the inner region of IC 348 (see text for the details). Among the most prominet features are the 11.2 μm PAH band and the molecular hydrogen lines at 12.28, 17.04 and 28.22 μm (zoomed in the next figure).

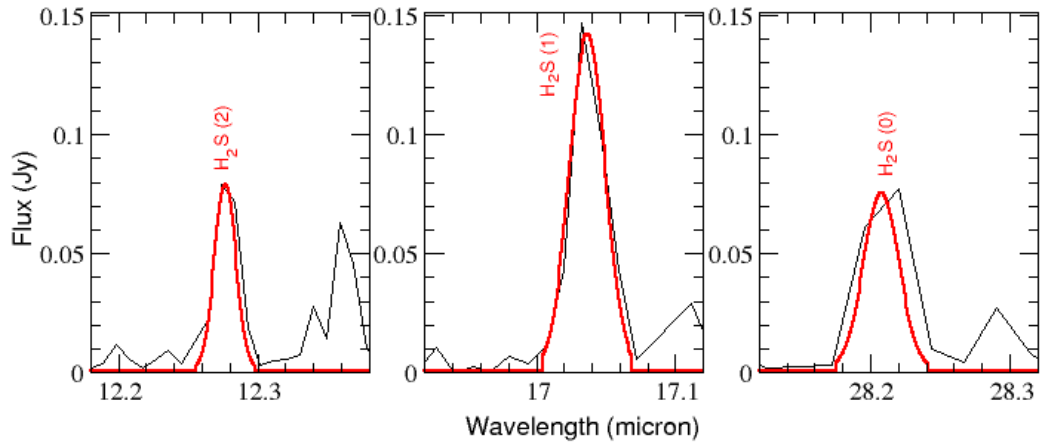

**Figure 3.** The three most intense H$_2$ transitions in the 10-30 μm region (S(0), S(1), S(2)) detected in the combined spectrum of the ISM in the IC 348 star-forming region. The central wavelengths provided by the fits show consistency within 0.01 μm with the laboratory wavelengths of these bands.

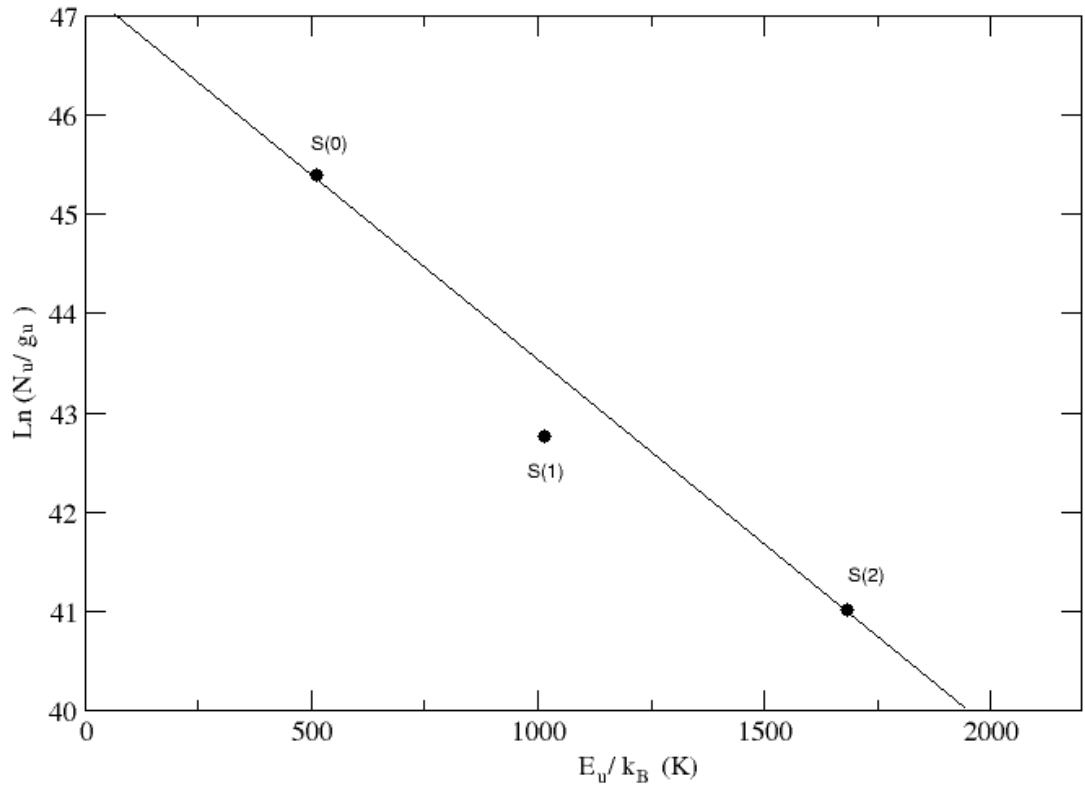

**Figure 4.** Excitaction diagram of $H_2$ for the gas in the core of the IC 348 star-forming region. The solid black line connects the two para-$H_2$ lines (S(0), S(2)). The slope corresponds to an excitation temperature of T=270 K.

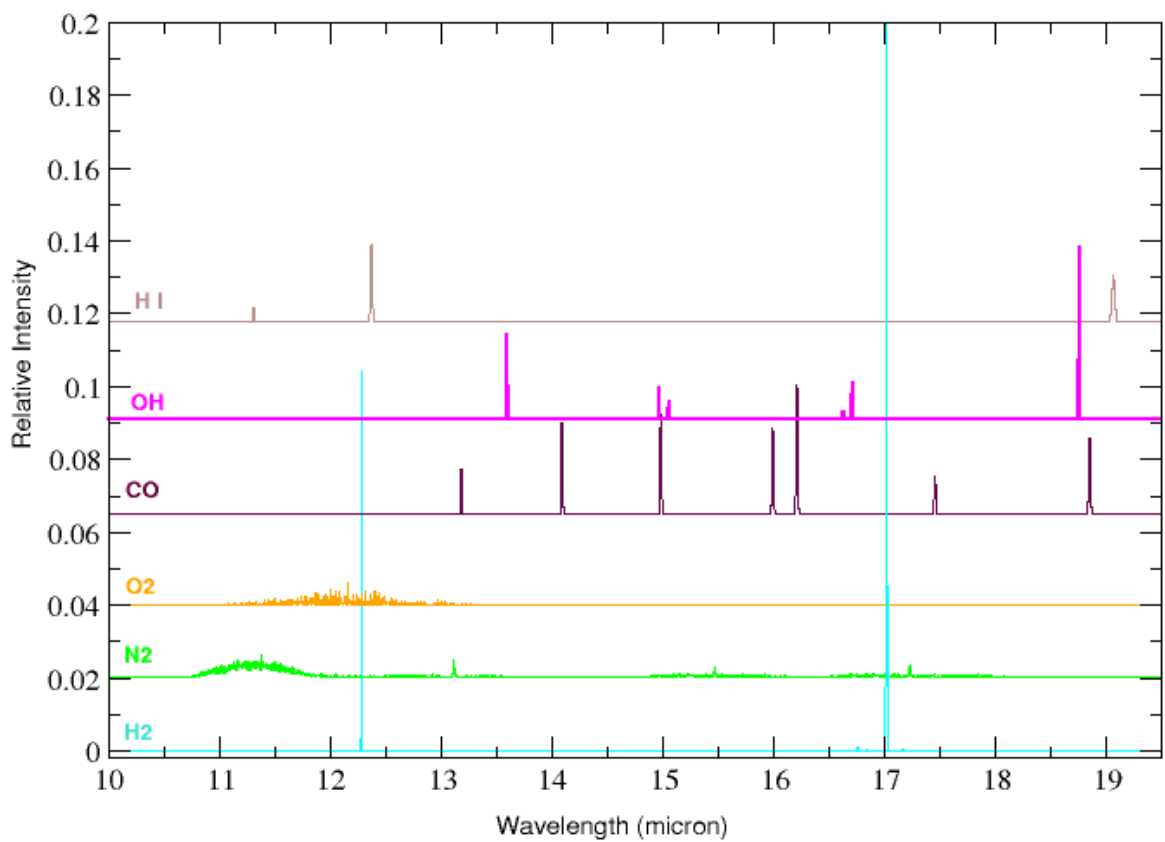

**Figure 5a.** HITRAN synthetic absorption spectra (240K and 0.001 atm) of several diatomic molecules and atomic hydrogen in the 10-19.5 μm spectral range

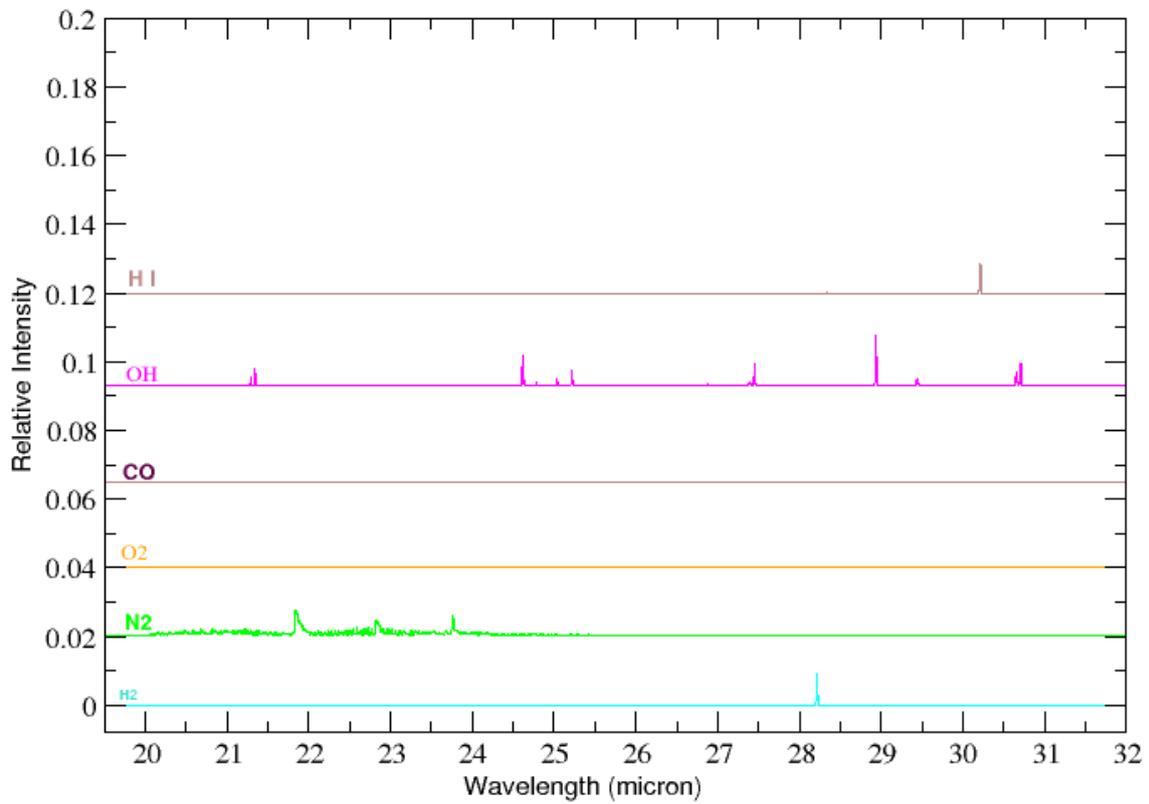

**Figure 5b.** HITRAN synthetic absorption spectra ( as in previous figure)

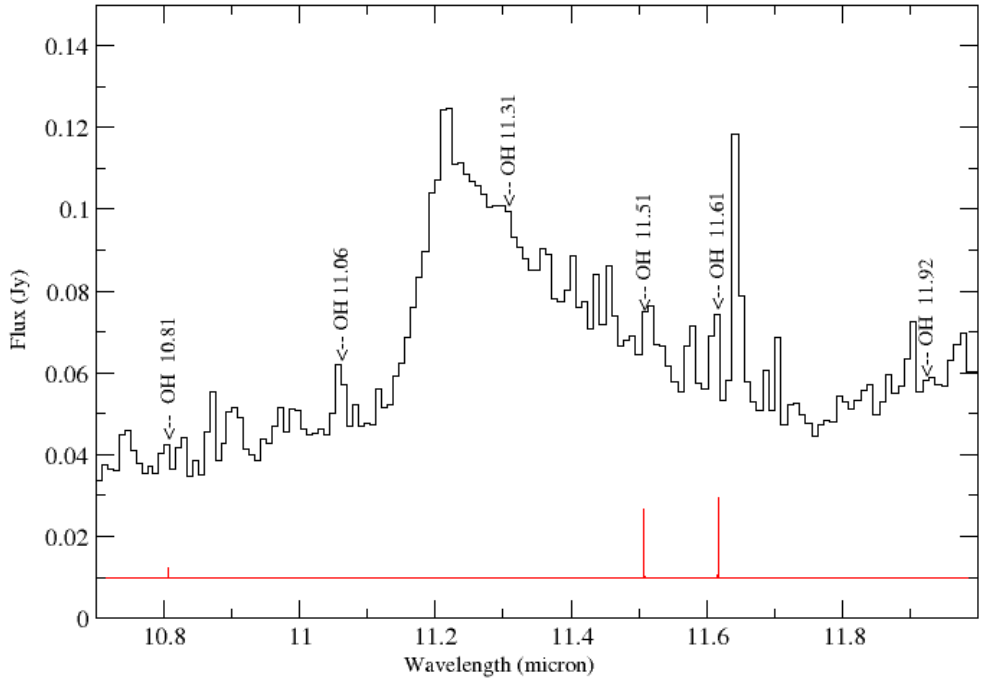

**Figure 6a.** Observed spectrum of IC 348 interstellar gas for the spectral range 10.7-12.0 μm (black line). OH transitions are marked. The OH spectral synthesis obtained with HITRAN for a temperature of 300 K is plotted at the bottom (red line). The lines at 10.81, 11.51 and 11.61 μm are observed with relative intensities similar to those in the synthetic spectrum. Additional OH bands associated with gas at much higher temperature (as explained in Najita et al. 2010) may be present in the spectrum at 11.06, 11.31 and 11.92 μm.

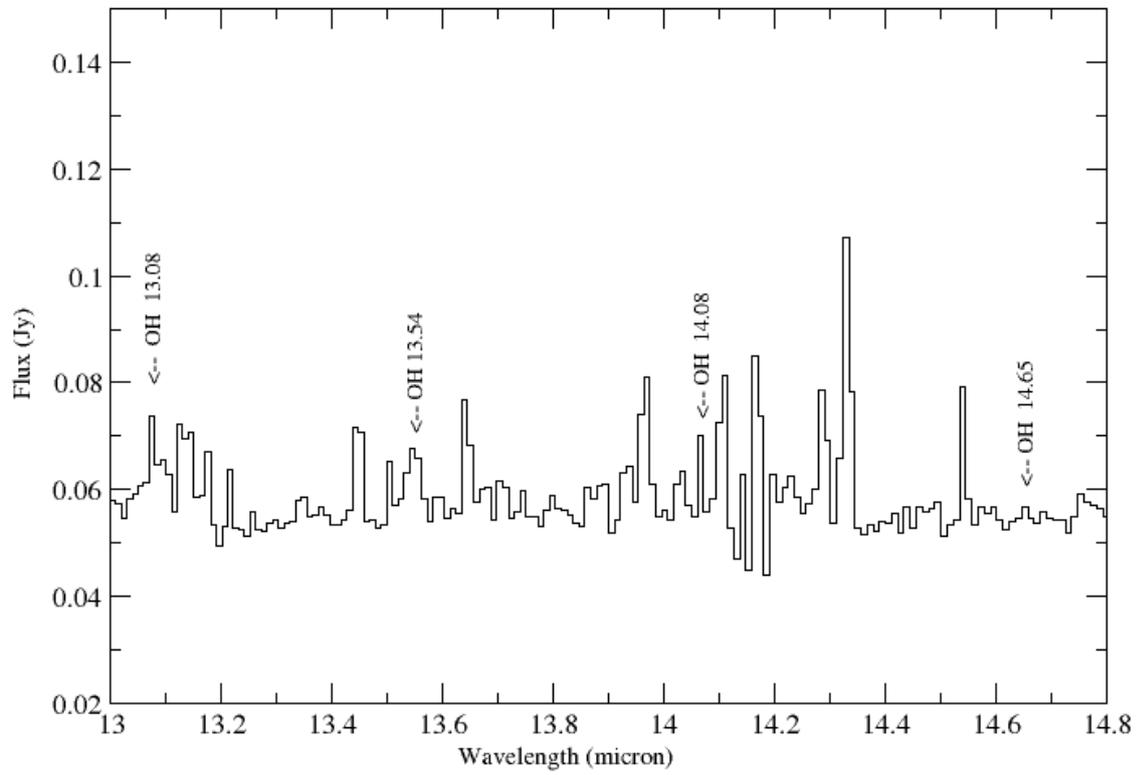

**Figure 6b.** Observed spectrum of IC 348 interstellar gas for the spectral range 13.0-14.8 μm with OH lines marked as in the previous figure.

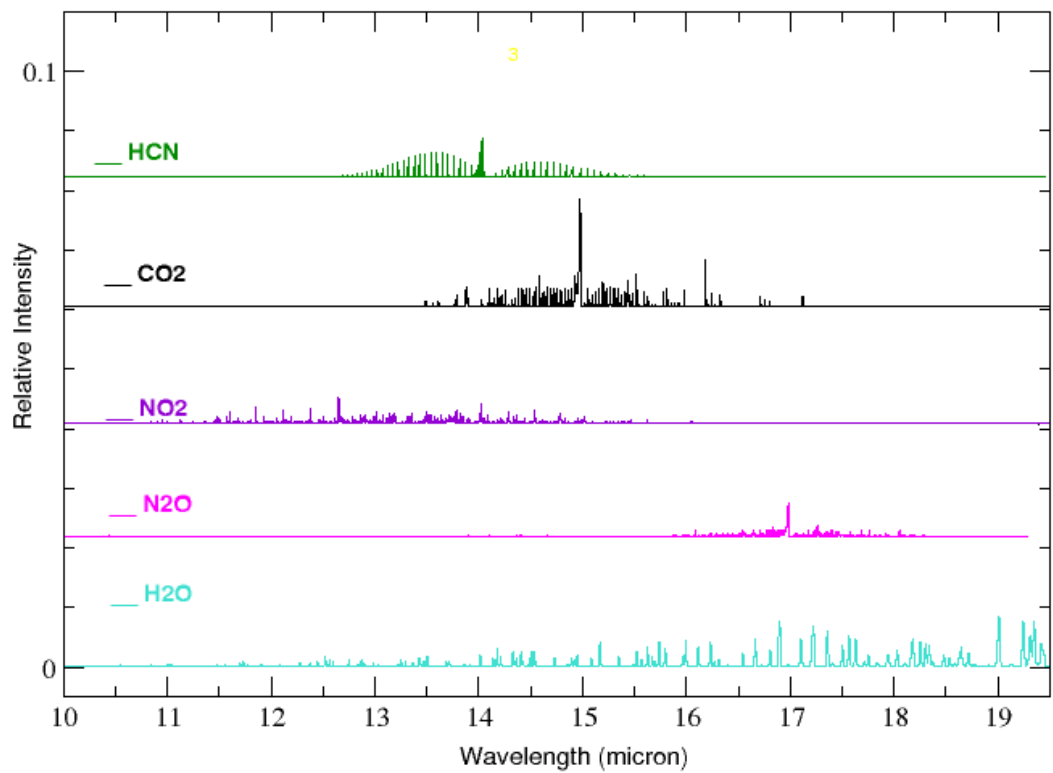

**Figure 7.** HITRAN based computations for molecules detected in diverse astrophysical environments: $H_2O$, $CO_2$, HCN, $NO_2$, $N_2O$.

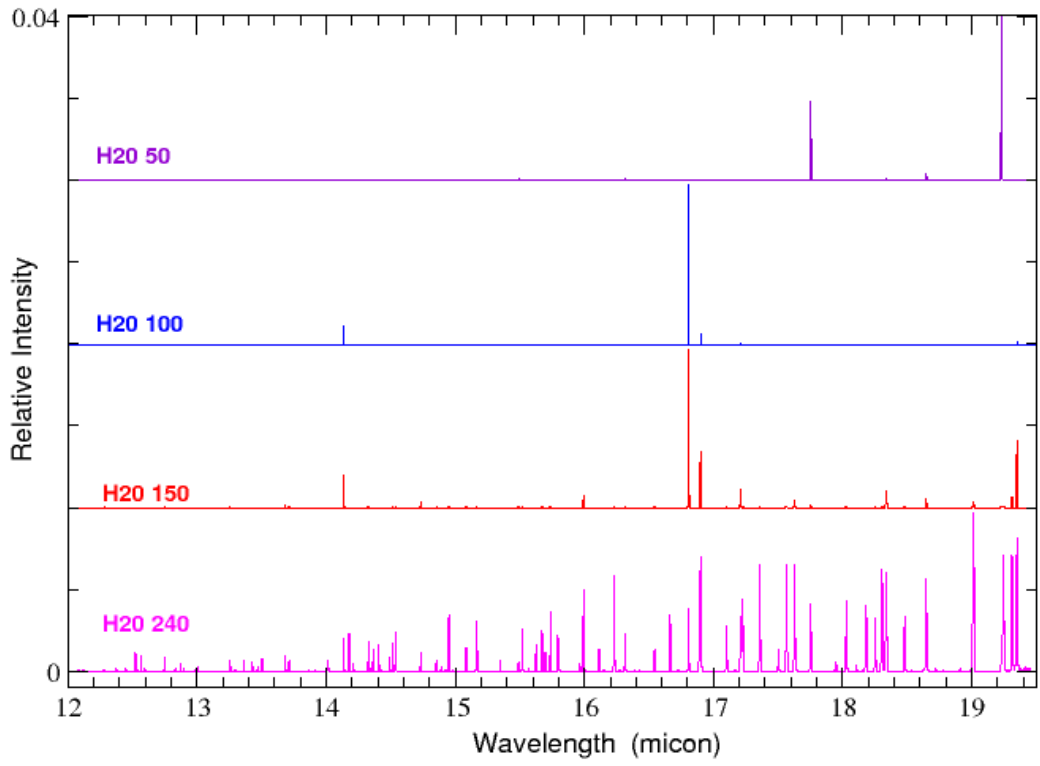

**Fig. 8a**  HITRAN synthetic water spectra at 50, 100, 150 and 240 K.

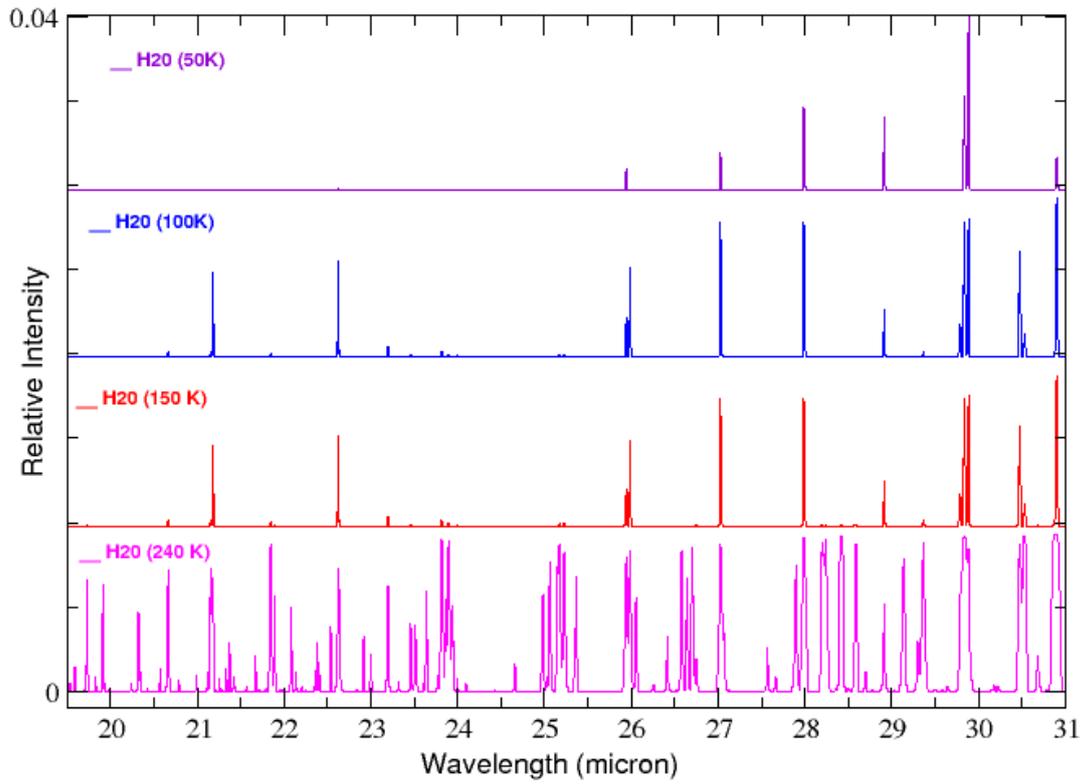

**Fig. 8b** Caption as in figure 8a for the spectral range 19.5-31.0 µm

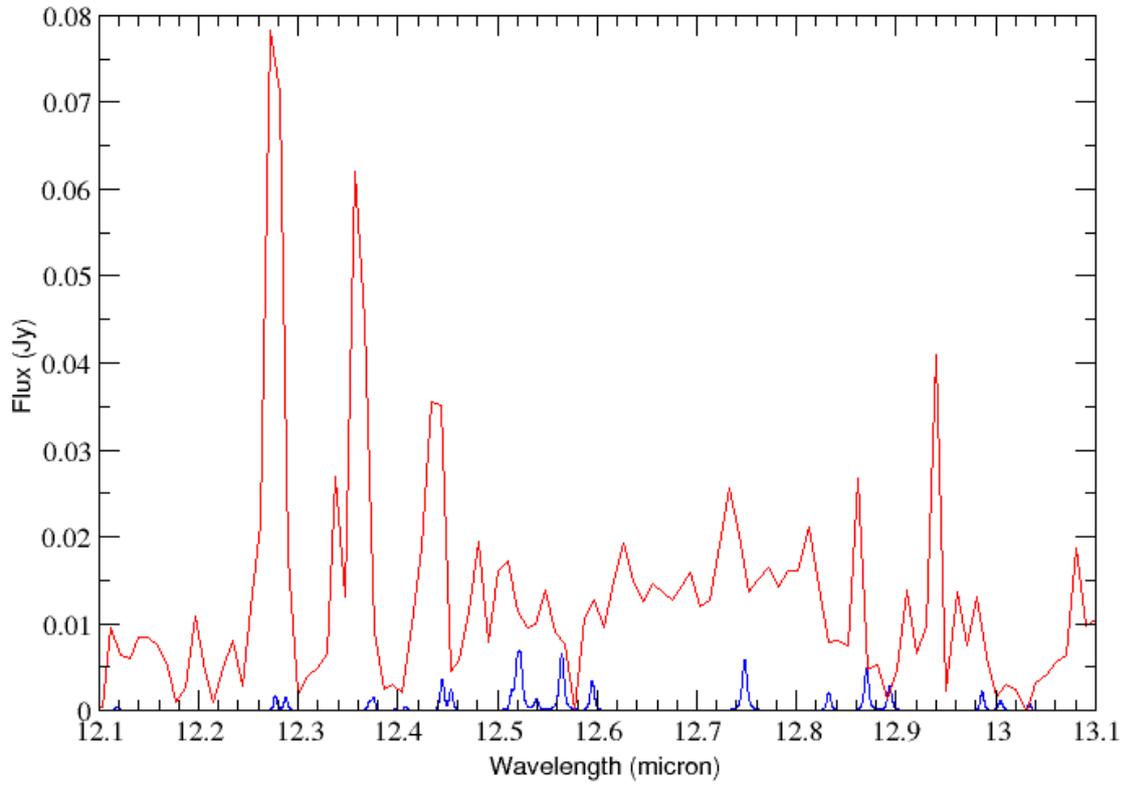

**Figure 9a.** Comparison of the observed IC 348 ISM spectrum (red solid line) with HITRAN spectral synthesis of $H_2O$ (240K , 0.001atm ).

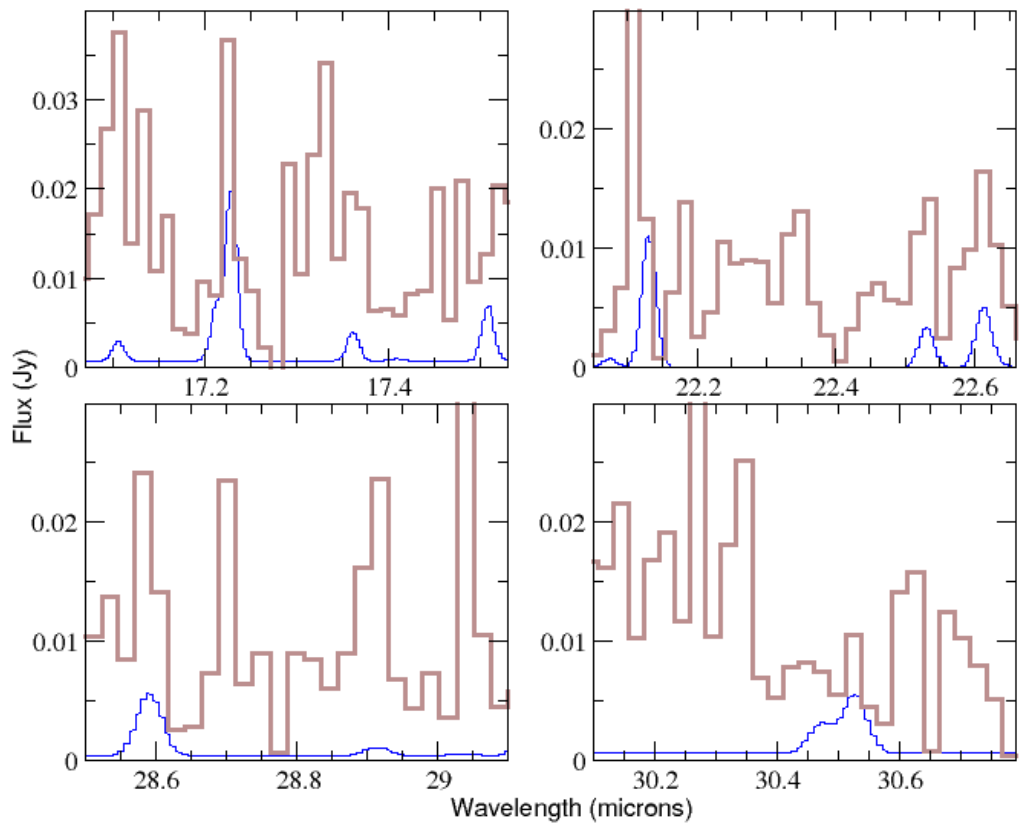

**Figure 9b.** The panels show several regions of the IC 348 gas spectrum (red line) and HITRAN (240K, 0.001atm) computations of $H_2O$ (blue line) at spectral resolution 600 to guide the identification of $H_2O$ lines in the observed spectrum.

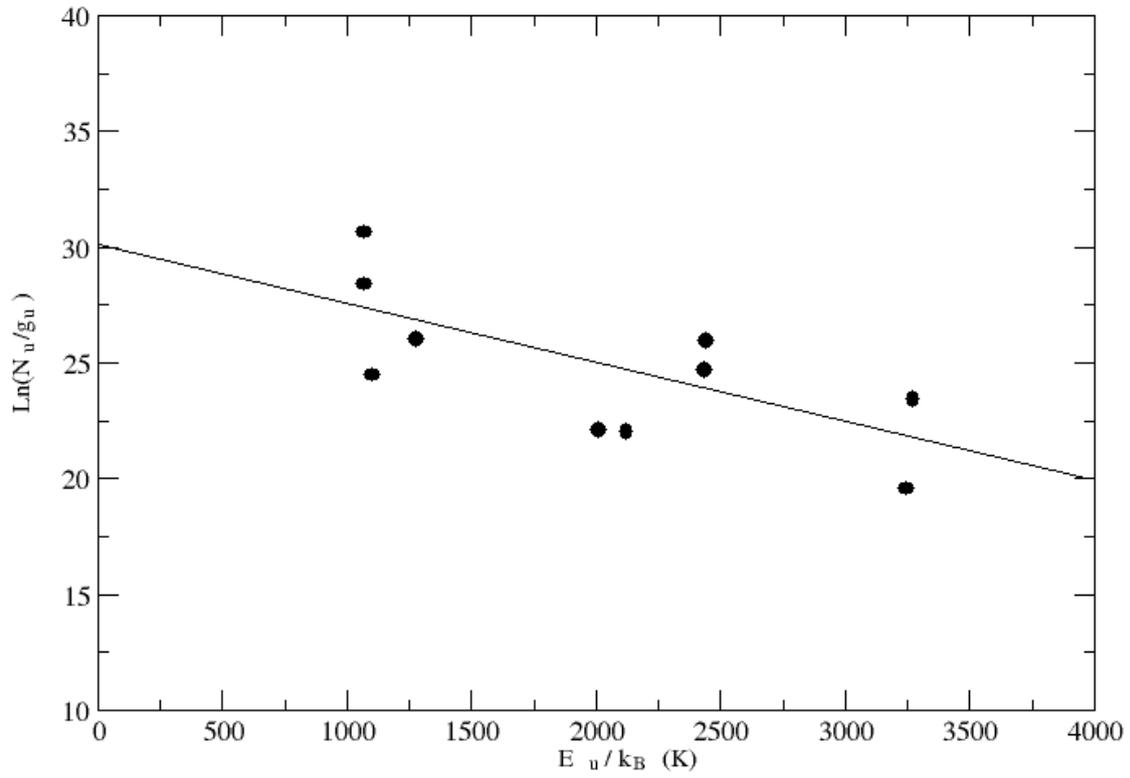

**Figure 9c.** Excitation diagram for water transitions detected in the ISM spectrum of IC 348. The solid line is a linear least squares fit to the data resulting in excitation temperature T = 385 K. The measurements error bars are smaller than the size of the symbols.

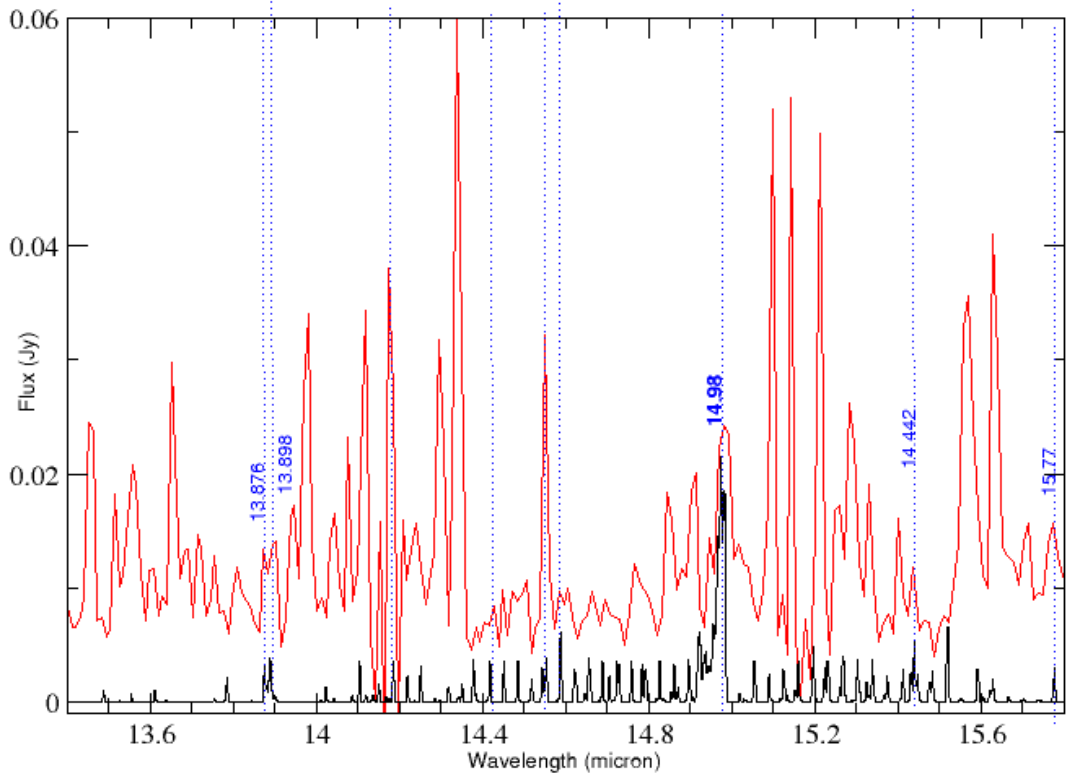

**Figure 10.** Comparison of the $CO_2$ synthetic HITRAN spectrum (solid black line) with the observed combined spectrum of the ISM in IC 348 (red line). The synthetic spectrum is scaled to match the strength of the 14.98 μm emission band. Other emission features in the spectrum which may display a significant contribution by $CO_2$ lines are marked.

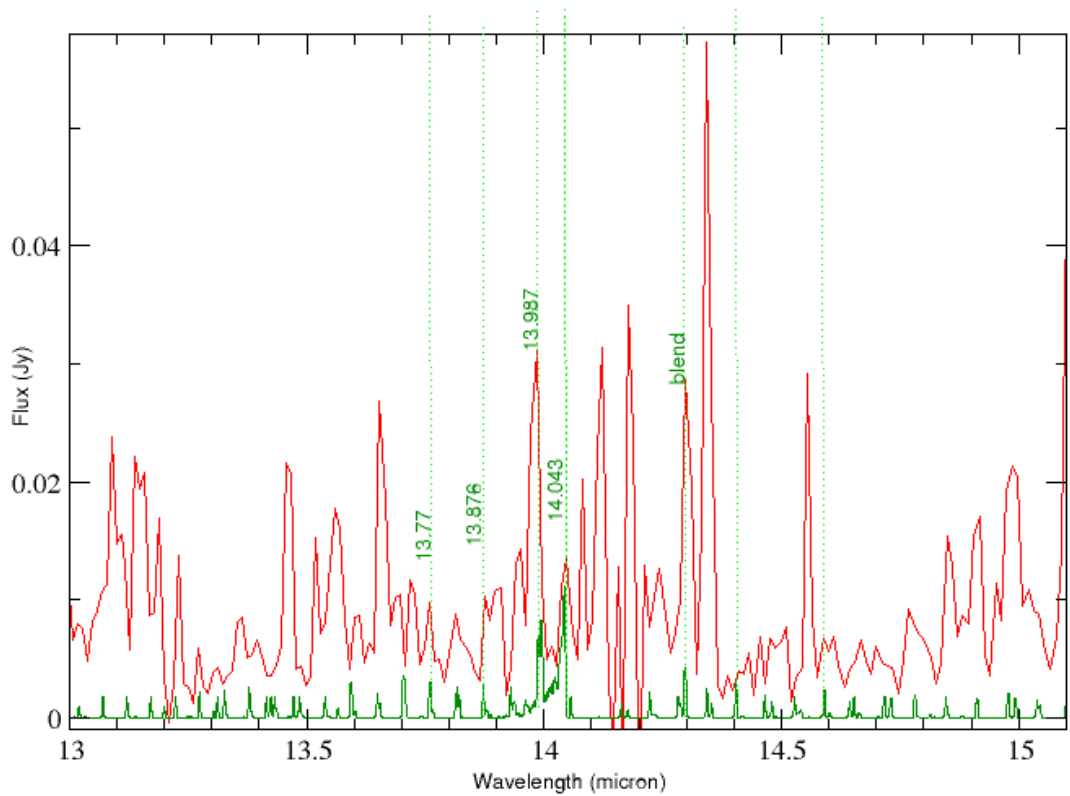

**Figure 11.** Comparison of HCN synthetic HITRAN spectrum with the observed ISM spectrum in IC 348. The synthetic spectrum (black line) is scaled to match the strength of the 14.04 μm emission feature which is ascribed to the HCN $v_2$ ro-vibrational band at 14.043 μm. Other observed emission features that could have a significant contribution from HCN are marked.

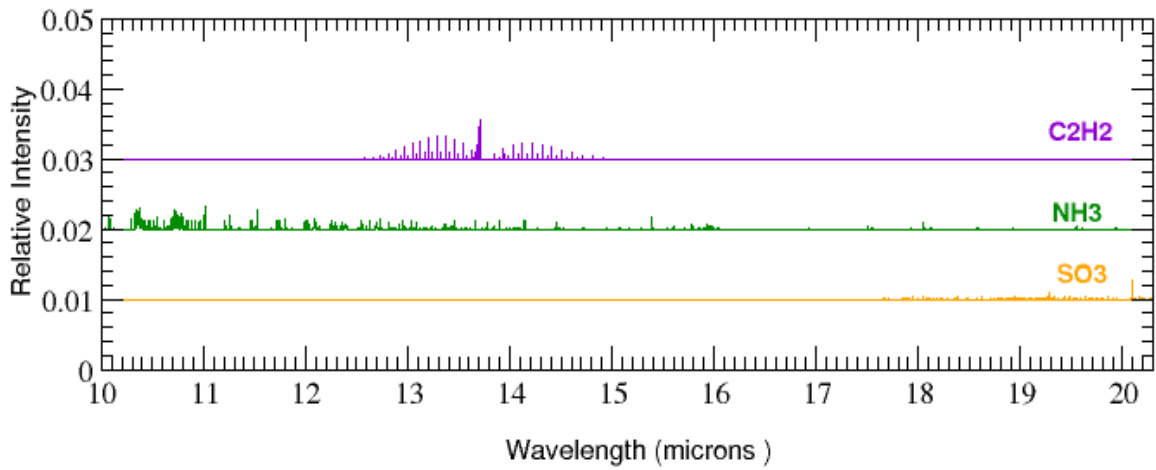

**Figure 12.** Synthetic HITRAN spectra for tetra-atomic molecules considered in this work.

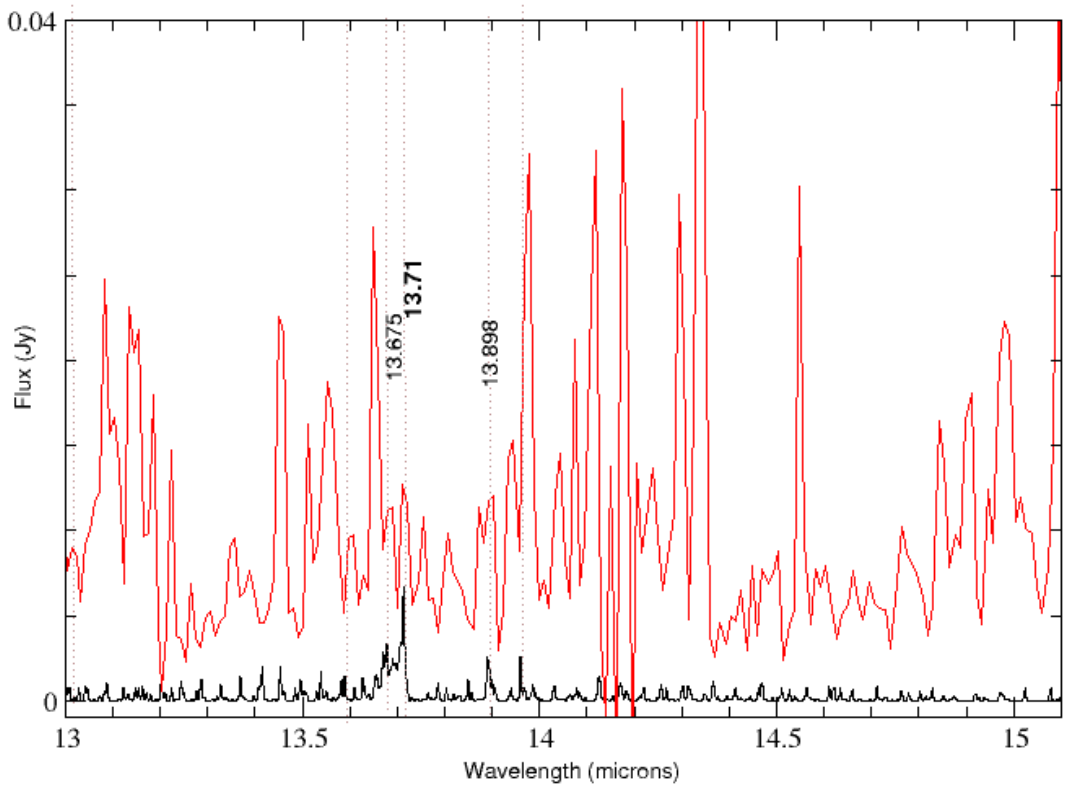

**Figure 13.** Comparison of $C_2H_2$ synthetic (HITRAN) spectrum with ISM observations in IC 348. The synthetic spectrum (black line) is scaled to match the emission feature at 13.71 μm and the strongest ro-vibrational band of $C_2H_2$. The emission lines at 13.675 and 13.898 μm could have a significant contribution from $C_2H_2$.

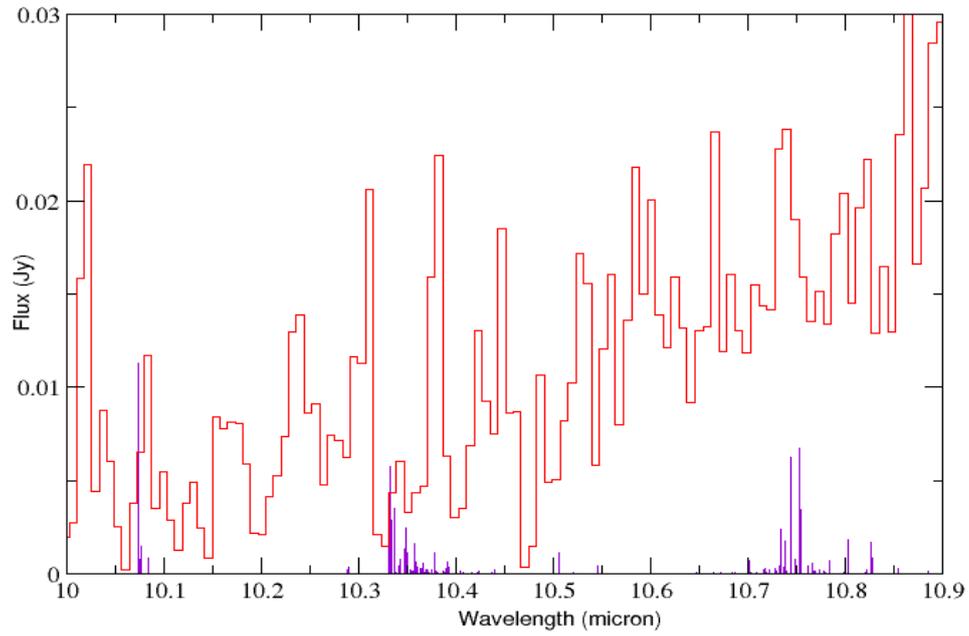

**Figure 14.** Comparison of NH3 synthetic (HITRAN) bands with the ISM observed IC 348 spectrum. The synthetic spectrum has been scaled to match the strength of the observed 10.74 μm feature. NH3 may also contribute to the weaker emission at 10.34 μm.

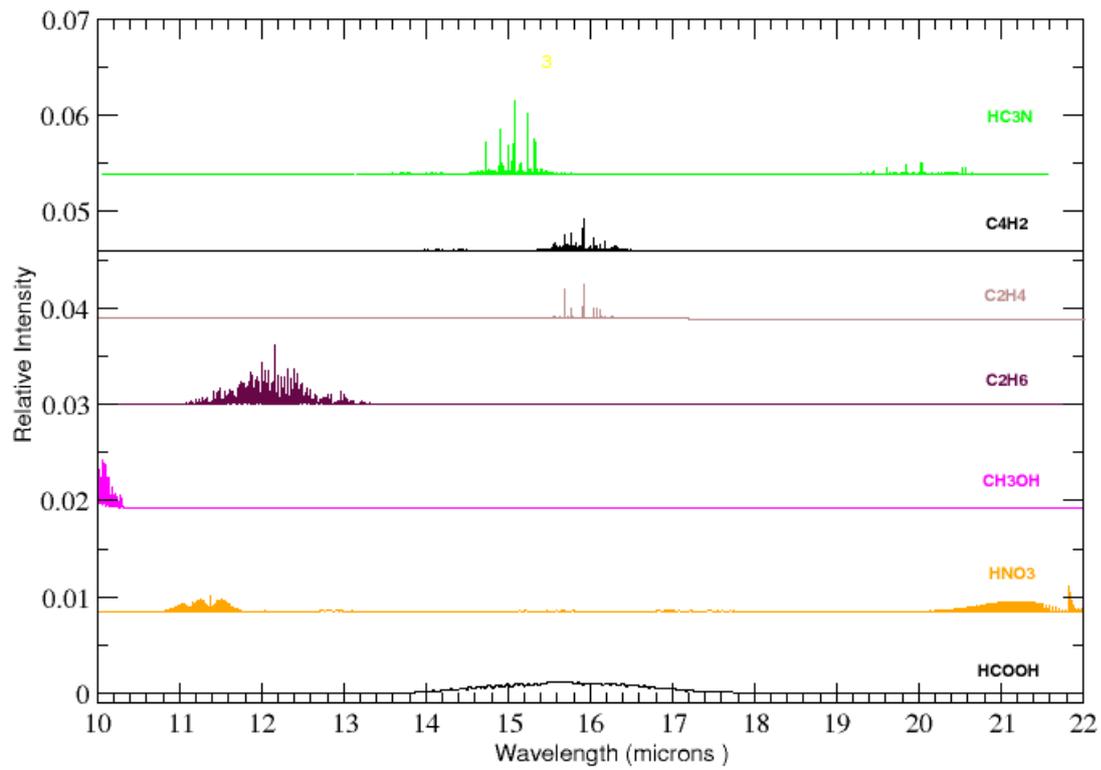

**Figure 15.** HITRAN synthetic spectra for polyatomic molecules at 240 K.

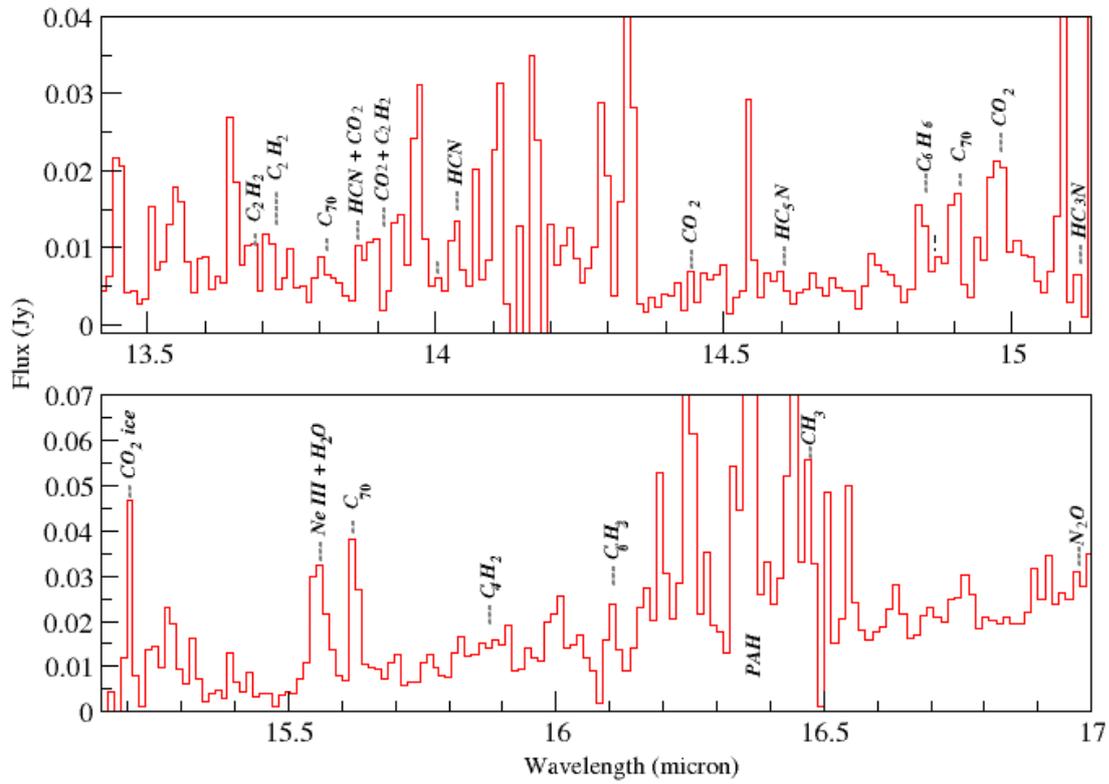

**Figure 16.** Probable identification of bands of several carbon bearing molecules (CO2, C2H2, C2H6 , C4H2  C6H2 , C6H6 , HCN , HC3N , HC5N, C70 ) in the spectrum of the ISM in IC 348.

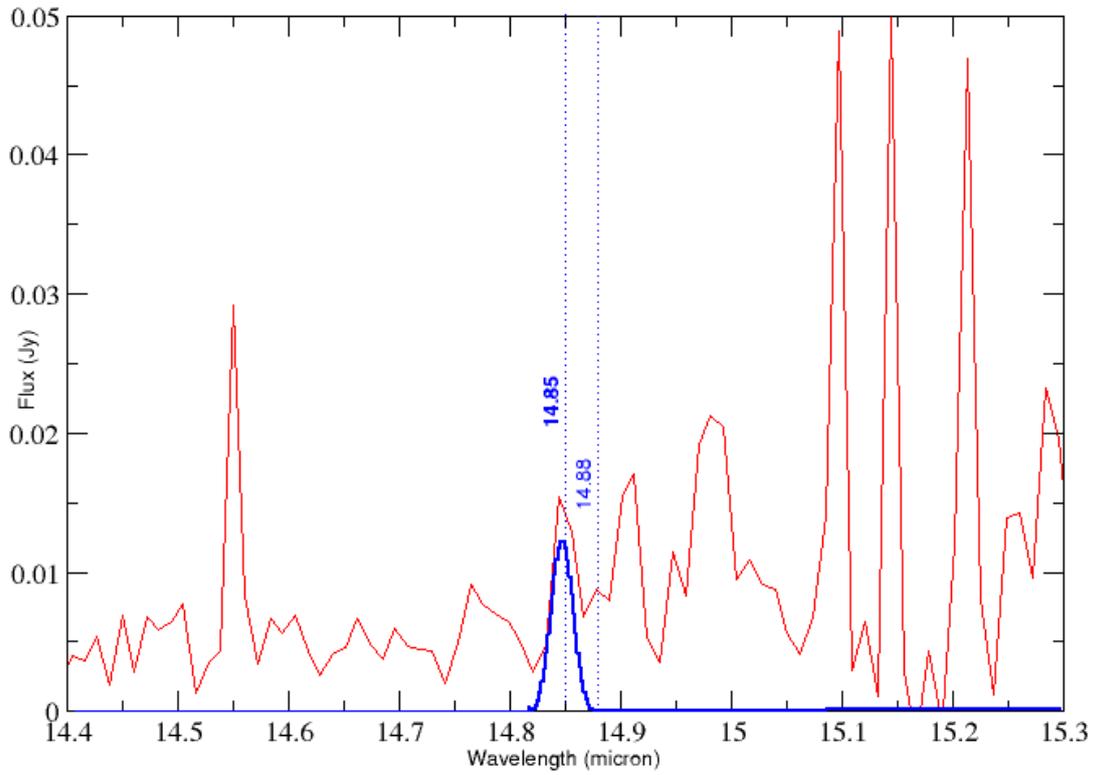

**Figure 17.** Fit to the emission feature ascribed to benzene ν4 Q-branch at 14.837 μm.

## TABLES

## Table 1. IC 348 ISM locations observed with Spitzer IRS

Short-High spectra

| AOR | R.A.$_{2000}$ | Dec$_{2000}$ |
|---|---|---|
| 22848000[c] | 03:44:38.90 | +32:10:03.0 |
| 22848512[c] | 03:44:41.66 | +32:06:45.4 |
| 22849024[c] | 03:45:00.87 | +32:09:15.5 |
| 22849536[c] | 03:44:22.88 | +32:04:57.0 |
| 22850048[c] | 03:44:42.71 | +32:03:29.8 |
| 22850560[c] | 03:44:36.08 | +32:00:14.7 |
| 22851072[c] | 03:43:50.10 | +32:08:17.9 |
| 22851584[c] | 03:44:33.23 | +31:59:54.1 |

Long-High spectra

| AOR | R.A.$_{2000}$ | Dec$_{2000}$ |
|---|---|---|
| 27542016_4[w] | 03:44:32.67 | +32:11:05.2 |
| 27542016_2[w] | 03:44:30.31 | +32:11:35.2 |
| 27542016_3[w] | 03:44:27.95 | +32:12:05.2 |
| 27542016_1[w] | 03:44:30.31 | +32:11:35.2 |

[c] program ID 40247 (N. Calvet, P.I.)
[w] program ID 50560 (D. Watson, P.I)

**Table 2. Measured fluxes of water lines in the spectrum of RNO 90**
(ID program: 50641; AOR 27061760)

| Wavelength (μm) | Flux[b] ($10^{-17}$ W m$^{-2}$) | Flux[c] ($10^{-17}$ W m$^{-2}$) |
|---|---|---|
| 15.16 + 15.17 | < 5.8 | 5.7 ± 0.9 |
| 15.57 + 15.62 | 18.0 ± 1.0 | 16.5 ± 1.0 |
| 15.74 | 7.3 ± 0.6 | 7.2 ± 0.6 |
| 17.10 | 9.0 ± 0.9 | 9.5 ± 0.9 |
| 17.23 | < 9.4 | 9.2 ± 0.9 |
| 22.54+22.62+22.64 | 12.5 ± 1.0 | 14.0 ± 1.0 |
| 23.46 + 23.51 | 12.8 | 13.0 ± 1.0 |
| 26.42 | < 2.0 | 2.2 ± 1.0 |
| 28.59 | 11.5 ± 0.6 | 9.0 ± 0.6 |
| 29.14 | < 4.6 | 3.9 ± 0.8 |
| 29.36 | < 0.8 | <1.0 |
| 30.47 +30.53 | 12.2 ± 0.7 | 13.5 ± 0.9 |
| 30.87 +30.90 | 10.5 ± 0.8 | 10.5 ± 0.8 |
| 31.74 +31.77 | < 2.6 | 2.5 ± 0.8 |
| 32.80 +30.83 | < 3.2 | 2.7 ± 0.8 |
| 32.92 +32.99+33.01 | 27.4 ± 0.8 | 25.0 ± 0.9 |

[b] Blevins et al (2016)
[c] This work (full aperture)

**Table 3. Molecular hydrogen in the gas at the core of IC 348**

| Transition | $\lambda$ [μm] | $E_u$ [K] | $A_{ul}$ [s$^{-1}$] | $g_u$ | $F$ [10$^{-18}$ Wm$^{-2}$] | $N_u$ [cm$^{-2}$] |
|---|---|---|---|---|---|---|
| H$_2$ v = 0–0 S(0) J = 2–0 | 28.219 | 509.9 | 2.943(–11) | 5 | 25±2 | 2.6E+20 |
| H$_2$ v = 0–0 S(1) J = 3–1 | 17.035 | 1015.1 | 4.761(–10) | 21 | 43±4 | 7.8E+19 |
| H$_2$ v = 0–0 S(2) J = 4–2 | 12.279 | 1681.9 | 2.755(–9) | 9 | 22±2 | 5.8E+18 |

Note: molecular parameters from Roueff et al. (2019) and HITRAN database

## Table 4. H2O in the gas of the core of IC 348

| Transit. | Wavelength (μm) | wavenum (cm$^{-1}$) | $A_{ij}$ (s$^{-1}$) | $E_{up}$ (K) | g | $F_{obs}$ (10$^{-18}$ Wm$^{-2}$) | $N_u$ (10$^{13}$cm$^{-2}$) | Notes |
|---|---|---|---|---|---|---|---|---|
| o-H$_2$O$^S$ 16$_{4\,13}$–15$_{1\,14}$ | **12.375** | 806.696 | 7.61 | 4948 | 105 | <1 | -- | u |
| p-H$_2$O$^S$ 11$_{3\,9}$–10$_{0\,10}$ | **12.407** | 805.996 | 4.19 | 4945 | 33 | <1 | -- | u |
| o-H$_2$O$^S$ 11$_{8\,3}$–10$_{5\,6}$ | **12.444** | 803.546 | 0.29 | 3629 | 69 | 15±2 | -- | str+b |
| o-H$_2$O$^S$ 13$_{7\,6}$–12$_{4\,9}$ | **12.453** | 802.989 | 1.04 | 4213 | 69 | <1 | -- | b |
| o-H$_2$O$^h$ 10$_{8\,2}$–9$_{5\,5}$ | **12.893** | 779.303 | 0.15 | 3217 | 21 | <1 | -- | u |
| o-H$_2$O$^B$ 12$_{3\,10}$–11$_{0\,11}$ | **15.738** | 635.397 | 1.10 | 2823 | 75 | <1 | -- | u |
| o-H$_2$O$^B$ 12$_{5\,8}$–11$_{2\,9}$ | **17.103** | 584.708 | 3.78 | 3273 | 75 | 5±1 | 0.11 | pr |
| p-H$_2$O$^B$ 11$_{3\,9}$–10$_{0\,10}$ | **17.225** | 580.536 | 0.97 | 2438 | 23 | 5±1 | 0.45 | r |
| o-H$_2$O$^B$ 11$_{2\,9}$–10$_{1\,10}$ | **17.358** | 576.114 | 0.96 | 2432 | 69 | 4±1 | 0.37 | r |
| o-H$_2$O$^B$ 10$_{8\,3}$–9$_{7\,2}$ | **22.538** | 443.695 | 33.15 | 3243 | 63 | 3±1 | 0.002 | b |
| o-H$_2$O$^B$ 6$_{5\,2}$–5$_{2\,3}$ | **22.619** | 442.088 | 0.06 | 1278 | 39 | 2±1 | 0.82 | b |
| o-H$_2$O$^B$ 5$_{5\,0}$–4$_{2\,3}$ | **22.639** | 441.714 | 0.014 | 1067 | 33 | 4±1 | 7.04 | s |
| p-H$_2$O$^B$ 5$_{5\,1}$–4$_{2\,2}$ | **23.458** | 426.293 | 0.015 | 1067 | 11 | 13±2 | 22.15 | vs |
| p-H$_2$O$^B$ 7$_{7\,1}$–6$_{6\,0}$ | **28.591** | 349.755 | 20.39 | 2006 | 15 | 4±1 | 0.006 | sr |
| p-H2O$^B$ 9$_{5\,5}$–8$_{4\,4}$ | **29.137** | 343.205 | 8.98 | 2122 | 19 | 2±0.5 | 0.007 | r |
| o-H$_2$O$^B$ 7$_{2\,5}$–6$_{1\,6}$ | **29.836** | 335.157 | 0.51 | 1100 | 45 | 3±1 | 0.19 | r |
| o-H$_2$O$^B$ 7$_{6\,1}$–6$_{5\,2}$ | **30.525** | 327.595 | 13.58 | 1749 | 45 | <1 | -- | u |
| o-H$_2$O$^B$ 7$_{6\,2}$–6$_{5\,1}$ | **30.529** | 327.557 | 13.57 | 1735 | 15 | <1 | -- | u |

$^B$ *Blevins et al 2016 ;$^h$ HiTRAN ; $^S$Salyk et al. 2011; s-strong, sr-strong resolved ,vs-very strong , b-blend, pr-partially resolved, r-resolved, u undetected*